\documentclass[12pt,letterpaper]{article}
\usepackage{jheppub}
\usepackage{graphicx}
\input{epsf}
\usepackage{epsfig}
\usepackage{epstopdf}
\usepackage{subfigure}
\usepackage{amsmath}
\newcommand{\be}{\begin{equation}}
\newcommand{\ee}{\end{equation}}

\DeclareMathOperator{\Tr}{Tr}

\usepackage{tikz}
\usetikzlibrary{calc}   
\usetikzlibrary{topaths}
\usepackage{pgfplots}
\newcommand{\labell}[1]{\label{#1}}

\newcommand{\bea}{\begin{eqnarray}}
\newcommand{\eea}{\end{eqnarray}}
\newcommand{\ba}{\begin{eqnarray}}
\newcommand{\ea}{\end{eqnarray}}

\newcommand{\beq}{\begin{equation}}
\newcommand{\eeq}{\end{equation}}
\newcommand{\beqa}{\begin{eqnarray}}
\newcommand{\eeqa}{\end{eqnarray}}
\newcommand{\beqar}{\begin{eqnarray*}}
\newcommand{\eeqar}{\end{eqnarray*}}

\newcommand{\reef}[1]{(\ref{#1})}
\newcommand{\ssc}{\scriptscriptstyle}
\newcommand{\eg}{{\it e.g.,}\ }
\newcommand{\ie}{{\it i.e.,}\ }
\newcommand{\comment}[1]{{\bf [[[#1]]]}}

\newcommand{\mt}[1]{\textrm{\tiny #1}}

\newcommand{\C}{\mathcal{C}}

\newcommand{\cO}{\mathcal{O}}

\newcommand{\lgb}{\lambda_{\text{\tiny{GB}}}}
\newcommand{\tL}{\tilde{L}}

\newcommand{\fin}{f_\infty}

\newcommand{\ta}{{\tilde \kappa}}

\newcommand{\te}{t_\mt{E}}

\newcommand{\F}{$F$}

\newcommand{\q}{a}
\newcommand{\qe}{\q_{\ssc E}}
\newcommand{\see}{S_{\ssc EE}}
\newcommand{\ka}{\kappa}
\newcommand{\kae}{\ka_{\ssc E}}
\newcommand{\hm}{\gamma} 
\newcommand{\cs}{c_{\ssc S}}
\newcommand{\cse}{c_{\ssc S,E}}
\newcommand{\ctt}{C_{\ssc T}}
\newcommand{\ctte}{C_{\ssc T,E}}
\newcommand{\lhat}{\hat\lambda}
\newcommand{\alhat}{\hat\alpha}

\renewcommand{\href}[2]{#2}

\title{Corner contributions to holographic entanglement entropy}
\author[a]{Pablo Bueno}
\author[b]{and Robert C. Myers}
\affiliation[a]{Instituto de F\'isica Te\'orica UAM/CSIC,\\C/ Nicol\'as Cabrera, 13-15, C.U. Cantoblanco, 28049 Madrid, Spain}
\affiliation[b]{Perimeter Institute for Theoretical Physics,\\31 Caroline Street North, Waterloo, Ontario N2L 2Y5, Canada}
\vspace{0.2cm}
\emailAdd{p.bueno@csic.es}
\emailAdd{rmyers@perimeterinstitute.ca}
\abstract{The entanglement entropy of three-dimensional conformal field theories contains a universal contribution coming from corners in the entangling surface. We study these contributions in a holographic framework and, in particular, we consider the effects of higher curvature interactions in the bulk gravity theory. We find that for all of our holographic models, the corner contribution is only modified by an overall factor but the functional dependence on the opening angle is not modified by the new gravitational interactions. We also compare the dependence of the corner term on the new gravitational couplings to that for a number of other physical quantities, and we show that the ratio of the corner contribution over the central charge appearing in the two-point function of the stress tensor is a universal function for all of the holographic theories studied here. Comparing this holographic result to the analogous functions for free CFT's, we find fairly good agreement across the full range of the opening angle. However, there is a precise match in the limit where the entangling surface becomes smooth, \ie the angle approaches $\pi$, and we conjecture the corresponding ratio is a universal constant for all three-dimensional conformal field theories. In this paper, we expand on the holographic calculations in our previous letter {\tt arXiv:1505.04804}, where this conjecture was first introduced.}

\preprint{IFT-UAM/CSIC-15-054
}

\arxivnumber{1505.07842}

\begin{document}
\maketitle

\section{Introduction}\labell{Introduction}

Entanglement entropy (EE) has emerged as a useful tool in a variety of research areas, including  condensed matter physics \cite{2003quant.ph..4098L,2005PhRvL..94f0503P,2008RvMP...80..517A,Grover:2012sp}, quantum information \cite{Nielsen,2008PhRvL.100g0502W}, quantum field theory (QFT) \cite{Casini:2009sr,Calabrese:2004eu,Klebanov:2007ws,Kitaev:2005dm,Casini:2004bw,Shiba:2013jja,Casini:2013rba} and quantum gravity \cite{Srednicki:1993im,Bombelli:1986rw,Callan:1994py,Bianchi:2012ev,Ryu:2006bv,Ryu:2006ef,Maldacena:2013xja,Nishioka:2006gr,Lewkowycz:2012mw,VanRaamsdonk:2009ar}.
In the context of quantum field theory, we define the EE for a spatial region $V$ as: $S=-\Tr \left( \rho_V \log \rho_V \right)$, where $\rho_V$ is the reduced density matrix computed by integrating out the degrees of freedom in the complementary region $\overline{V}$. The focus of the discussion in this paper comes from considering the EE for a three-dimensional conformal field theory (CFT), which will have an expansion of the form
\be
S_{\ssc EE}=c_{1} \,\frac{\cal A }{\delta}-\q\, \log{\left(H/\delta\right)}-2\pi c_0+\mathcal{O}\left(\delta/H \right)\, , \labell{one}
\ee 
where $\cal A$, $H$ and $\delta$ are, respectively, the perimeter of the entangling surface,  some macroscopic length characteristic of the geometry (\eg we could choose $H=\cal A$) and a short-distance cut-off needed to regulate the calculation. Of course, the first term in this expansion is the celebrated `area law' contribution to the EE \cite{Bombelli:1986rw,Srednicki:1993im}. However, the dimensionless coefficient $c_1$ of this linear divergence depends on the details of the regulator and so cannot be used to characterize the underlying CFT. In contrast, in the absence of the logarithmic term (see below), the constant $c_0$ is a universal constant intrinsic to the CFT\footnote{Of course, in gapped systems with topological order, this finite contribution would correspond to the topological entanglement entropy \cite{2006PhRvL..96k0405L,Kitaev:2005dm,2005PhLA..337...22H}.} and also the geometry of the {\it smooth} entangling surface (the boundary of the region $V$). For example, when the latter is a circle, $c_0$ plays the role of a `central charge' in the $F$-theorem \cite{Myers:2010xs,Myers:2010tj,Fthem1,Fthem2,proof}. 

Another universal contribution in eq.~\reef{one}  is the one proportional to $\log(H/\delta)$, which arises when the entangling surface  contains corners \cite{Casini:2006hu,Casini:2008as,Hirata:2006jx,Fradkin:2006mb}\footnote{Our discussion focuses on three-dimensional CFT's, however, similar logarithmic contributions may appear in theories which break conformal symmetry \cite{Hung:2011ta,Huijse:2011ef,Ogawa:2011bz,Dong:2012se,Bueno:2014oua}.}${}^,$\footnote{The generalization to higher-dimensional singular surfaces was examined in \cite{Myers:2012vs}.} --- see figure \ref{une}.  Hence the dimensionless coefficient $\q$ is a function of the opening angle, \ie $\q=\q(\Omega)$. In our discussion, we focus on the contribution of a single corner in the entangling surface. If several corners were present, the coefficient of logarithmic contribution to the EE would simply involve the sum of independent contributions $\q(\Omega_i)$ where $\Omega_i$ is the opening angle of the $i$'th corner. The form of the function $\q(\Omega)$ is constrained by various properties of entanglement entropy \cite{Casini:2006hu,Casini:2008as,Hirata:2006jx}:  for pure states, the fact that $\see(V)=\see(\overline{V})$ requires that $\q(\Omega)=\q(2\pi-\Omega)$. Further, strong subadditivity and Lorentz invariance impose
 \beq
\q(\Omega)\geq 0\,,\qquad \partial_\Omega\q(\Omega)\leq 0\qquad{\rm and}\qquad \partial^2_\Omega\q(\Omega)\ge \frac{|\partial_\Omega\q(\Omega)|}{\sin\Omega}
\qquad{\rm for} \qquad \Omega\leq \pi\,, \labell{restrict} 
\eeq
\ie $\q(\Omega)$ is a positive convex function on the range $0\le\Omega\le\pi$. 
\begin{figure}[h]\hspace{4.7cm}
   \includegraphics[scale=0.2]{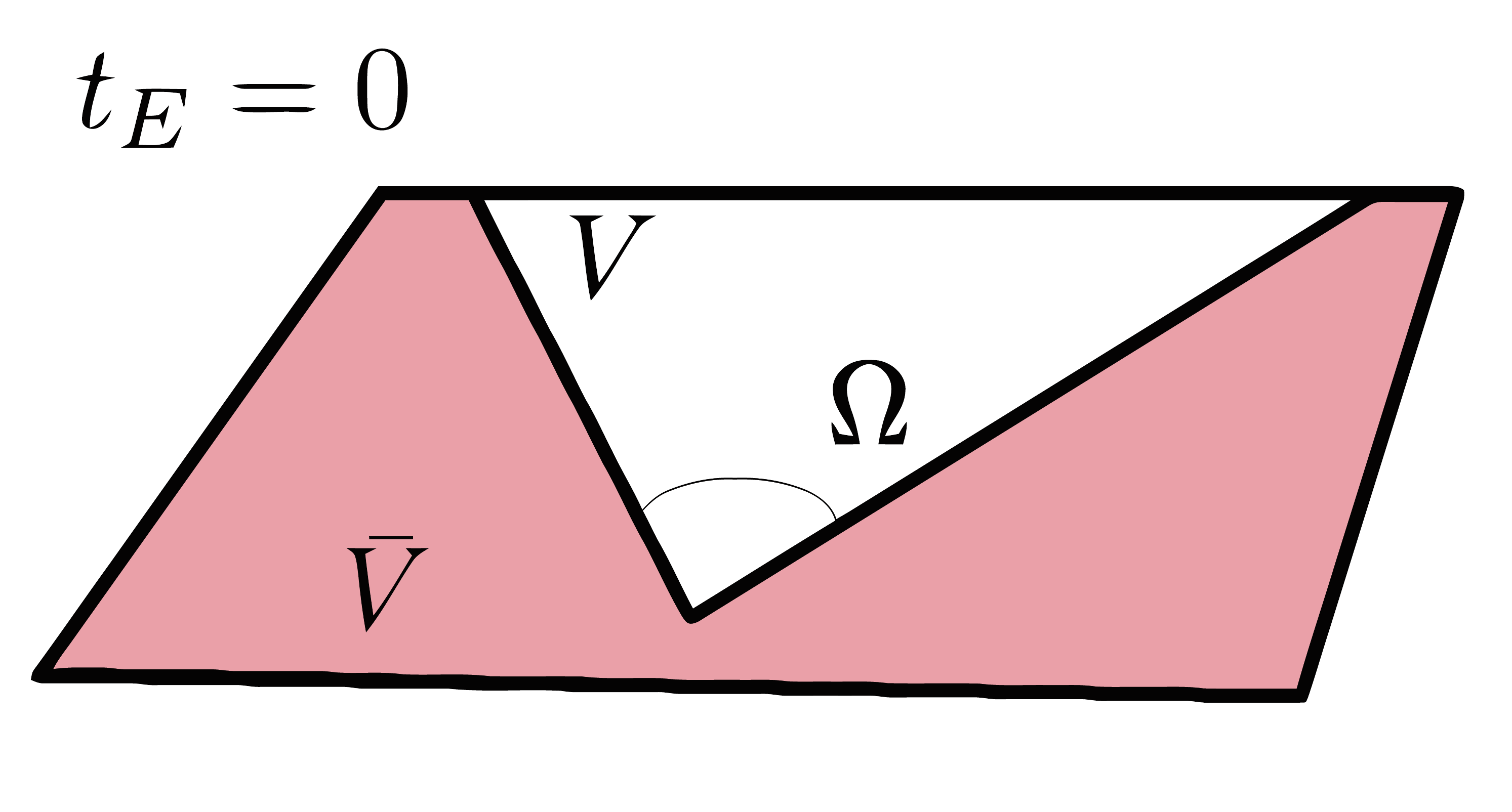}
\caption{(Colour online) A corner in the entangling surface with opening angle $\Omega$.}
\labell{une}
\centering
\end{figure}

In fact, the functional form of $\q(\Omega)$ is precisely constrained in particular limits. For small opening angles, the function has a pole with
\beq
\lim_{\Omega\to0} \q(\Omega)\equiv \frac{\kappa}{\Omega}+\cdots\,.
\labell{ka}
\eeq
As we will review in appendix \ref{app4}, this form for small angles can be fixed by using a conformal mapping to relate the universal corner contribution to the universal contribution for a narrow strip. Of course, $\q(\Omega)$ vanishes when the entangling surface becomes smooth, \ie $\q(\pi)=0$. Further, we can expect that $\q(\Omega)$ is smooth in the vicinity of $\Omega=\pi$ and hence the constraint $\q(\Omega)=\q(2\pi-\Omega)$ (for pure states) requires that to leading order,
\be\labell{sigmA}
\q(\Omega\sim \pi)\simeq  \sigma\, (\pi-\Omega)^2 +\cdots\,.
\ee
In fact, this constraint requires that $\q(\Omega)$ can be represented in a Taylor series with only even powers of $(\pi-\Omega)$ \cite{Casini:2006hu}. Hence we may use $\q(\Omega)$ in the limits $\Omega\to0$ and $\Omega\to\pi$ to define two interesting coefficients, $\kappa$ and $\sigma$, which characterize the underlying CFT.

The corner contribution to the entanglement entropy has been studied in a variety of systems: free scalar and fermion field theories \cite{Casini:2006hu,Casini:2009sr,Casini:2008as}, calculations at a quantum critical point \cite{2004AnPhy.310..493A}, numerical simulations in interacting lattice models \cite{PhysRevB.84.165134,PhysRevLett.110.135702,PhysRevB.86.075106,2013NJPh...15g3048I}, interacting scalar field theories \cite{2014arXiv1401.3504K} and also holographic calculations with Einstein gravity in the bulk \cite{Hirata:2006jx}. The results obtained in the literature suggest that $\q(\Omega)$ contains interesting and unambiguous information about the underlying quantum field theory. In particular, it appears to be an interesting measure of the number of degrees of freedom --- see, \eg \cite{2014arXiv1401.3504K,Casini:2006hu,Casini:2008as}. By the latter proposition, we would expect that the coefficients, $\kappa$ and $\sigma$, will themselves characterize the number of degrees of freedom in the underlying CFT.\footnote{Refs.~\cite{Casini:2009sr,Casini:2008as} discussed $\sigma$ for this purpose in the context of free field theories.} Motivated by this idea, we will take the liberty to refer to these coefficients as `central charges,' in a certain abuse of notation. 

In this paper,  we will study the universal term arising from the presence of corners in the entangling surface for three-dimensional holographic conformal field theories. 
One of our objectives is to study if the corner charges above have any simple relation to any other known constants, which provide a similar counting of degrees of freedom and might be accessed with more conventional probes of the theory, or if $\kappa$ and $\sigma$ are really distinct quantities. As we will discuss below, we can not make a meaningful comparison if the bulk theory corresponds to Einstein gravity. Hence our approach will be to study the corner contributions for a family of extended holographic models which include higher curvature interactions in the bulk gravity theory. Generally, any quantities in the corresponding dual boundary theories, \eg the corner term, will now depend on the new (dimensionless) gravitational couplings for these higher order terms. This additional dependence on the new couplings allows us to make a nontrivial comparison of $\kappa$ and $\sigma$ with various other constants in the boundary CFT's. In particular, we will compare with the coefficients appearing in the universal terms in the EE of a strip and of a disk, in the thermal entropy density, and in the two-point function of the holographic stress tensor.

In fact, beyond the corner charges, the entire functional form of $\q(\Omega)$ is characteristic of the underlying CFT. Hence another interesting question to consider is how this function changes with the inclusion of higher curvature interactions in the bulk. In this case, we find that for all of the holographic models studied here, $\q(\Omega)$ is only modified by an overall factor but the functional dependence on $\Omega$ is not modified by the new gravitational interactions.  However, as discussed in section \ref{goliger}, we do not believe that this behaviour is universal and that the functional form of $\q(\Omega)$ will be modified with sufficiently general higher curvature theories in the bulk.
One simple consequence of $\q(\Omega)$ not being changed here is that the two corner charges are simply related in all of our holographic models, \ie we will see that $\kappa/\sigma=4\,\Gamma(3/4)^4$. Hence we focus most of our discussions on the small angle charge $\kappa$ in the following. 

A final question, which we consider below, is whether our holographic analysis can reveal any features of the corner contribution which are universal to all three-dimensional CFT's. This question, which we examine in section \ref{free}, was originally addressed in our previous letter \cite{BuenoMyersWill}. Here we compare our holographic results with the corner terms in the free CFT's consisting of a conformally coupled massless scalar and of a massless fermion, as were calculated in \cite{Casini:2006hu,Casini:2009sr,Casini:2008as}.\\

\noindent Let us now summarize our key results:

The results for the ratios of the corner charge $\kappa$ with other various coefficients in the dual boundary theory are given in Table \ref{table}. The most interesting ratio is $\kappa/\ctt$, the corner charge over the central charge appearing in two-point function of the stress tensor \reef{t2p}, which is independent of all of the gravitational couplings.  Hence this ratio is universal for the broad class of holographic CFT's studied here.

In fact, as we noted above, the functional form of $\q(\Omega)$ is not modified by any of the higher curvature interactions, except for an overall factor. Given the above result, the entire function $\q(\Omega)/\ctt$ is universal for the broad class of holographic CFT's studied here. This holographic result suggests that this normalization provides an interesting way to compare the corner contribution between any general three-dimensional CFT's. Comparing our holographic result with the corresponding free field results,\footnote{Similar comparisons were made in \cite{Casini:2008as}, but without normalizing by the central charge $\ctt$.} we see that the free field curves agree with the holographic result remarkably well --- see figure \ref{fiqct}. The free fermion and scalar curves deviate from the holographic result by less than $2.5\%$ and $13\%$, respectively. Hence we suggest that the holographic expression for $\q(\Omega)/\ctt$, which is easily evaluated across the full range of $\Omega=0$ to $\pi$, provides a good benchmark with which to compare the analogous results for general three-dimensional CFT's. 

The maximum discrepancy between the holographic and free field results for $\q(\Omega)/\ctt$ occurs as $\Omega\to0$ but somewhat surprisingly they agree perfectly in the limit $\Omega\to \pi$, as first stated in \cite{BuenoMyersWill}. That is, the holographic CFT's and the two free field theories exhibit the same ratio 
\be \label{unisc2}
\frac{\sigma}{\ctt}=\frac{\pi^2}{24}\,.
\ee
This remarkable result leads us to conjecture that this ratio is in fact a universal constant for general conformal field theories in three dimensions. \\

The remainder of the paper is organized as follows: In section \ref{seckink}, we first review the holographic calculation of the entanglement entropy for a corner in the boundary of AdS$_4$ with Einstein gravity in the bulk. Then in section \ref{hcg}, we study the effects of adding various higher curvature interactions to the bulk gravity theory on the universal corner term. In doing so, we show that the functional form of $\qe(\Omega)$ is universal to all of the theories considered here and evaluate the small angle charge $\kappa$ appearing in each case. In section \ref{witncc}, we compare this corner charge in the higher curvature theories with similar quantities appearing in other physical observables, \ie the coefficients appearing in the universal contribution in the entanglement entropy of a strip and of a disk, in the thermal entropy density and in the two-point correlator of the stress tensor.  In section \ref{discuss}, we summarize our results. We also discuss the possibility of modifying the shape of the extremal surface in the holographic entanglement entropy in more general higher curvature theories of gravity, and hence modifying the functional form of $\q(\Omega)$ in the dual boundary theories. We also comment on the relation between our holographic results and the analogous results obtained for free field theories. In appendix \ref{app1}, we explain our conventions and notation in the calculations in section \ref{seckink}. In appendix \ref{app4}, we explain the conformal mapping which relates the corner charge $\ka$ with the coefficient of the universal term in the entanglement entropy of a strip. In appendix \ref{appf}, we compute the corner contribution for a general $f(R)$ theory and explain in some detail the linearized equations of motion used to compute the two-point function of the stress tensor. Finally in appendix \ref{integrals}, we present the integrals used in \cite{Casini:2009sr,Casini:2006hu,Casini:2008as} to evaluate the coefficient $\sigma$ for the free massless scalar and fermion theories and show that when evaluated with sufficient precision that they yield the simple rational values predicted by our conjecture \reef{unisc2}.


\section{Corner term in holographic entanglement entropy}\labell{seckink}

In this section we study the corner contribution to the entanglement entropy for holographic CFT's dual to higher curvature theories of gravity. In particular, we will consider bulk actions which contain general curvature-squared interactions and which are functions of Lovelock densities \cite{Sarkar}. However, we begin by reviewing the calculation of the corner contribution to holographic entanglement entropy with just Einstein gravity in the bulk, which was originally performed in \cite{Hirata:2006jx}.

The bulk geometry will be four-dimensional Euclidean anti-de Sitter space in Poincar\'e coordinates\footnote{See appendix \ref{app1} for conventions.}
\begin{equation}
ds^2=\frac{\tilde{L}^2}{z^2}\left(dz^2+d\te^2 +d\rho^2+\rho^2d\theta^2  \right)\, ,\labell{ads4}
\end{equation}
which is a solution for Einstein gravity coupled to a negative cosmological constant
\begin{equation}
I_0=\frac{1}{16\pi G}\int d^4x\sqrt{g}\left[\,\frac{6}{L^2}+R\, \right]  \labell{Einact}
\end{equation}
as long as we set $\tL=L$. The dual boundary theory then lives in the flat three-dimensional geometry with metric $d\tilde{s}^2=d\te^2 +d\rho^2+\rho^2d\theta^2$. The region for which we calculate the entanglement entropy will be defined as $V= \left\{\te=0,\rho>0,|\theta|\le  \Omega/2  \right\}$, as illustrated in figure \ref{une}. Hence the entangling surface $\partial V$ has a corner with opening angle $\Omega$ at the origin. Note that in the following, at as well as the usual short-distance cut-off $\delta$, we will also introduce an infrared regulator scale, \ie $\rho_{\ssc max}=H$, to ensure that the entanglement entropy does not diverge.

\begin{figure}[h]\hspace{3cm}
   \includegraphics[scale=0.35]{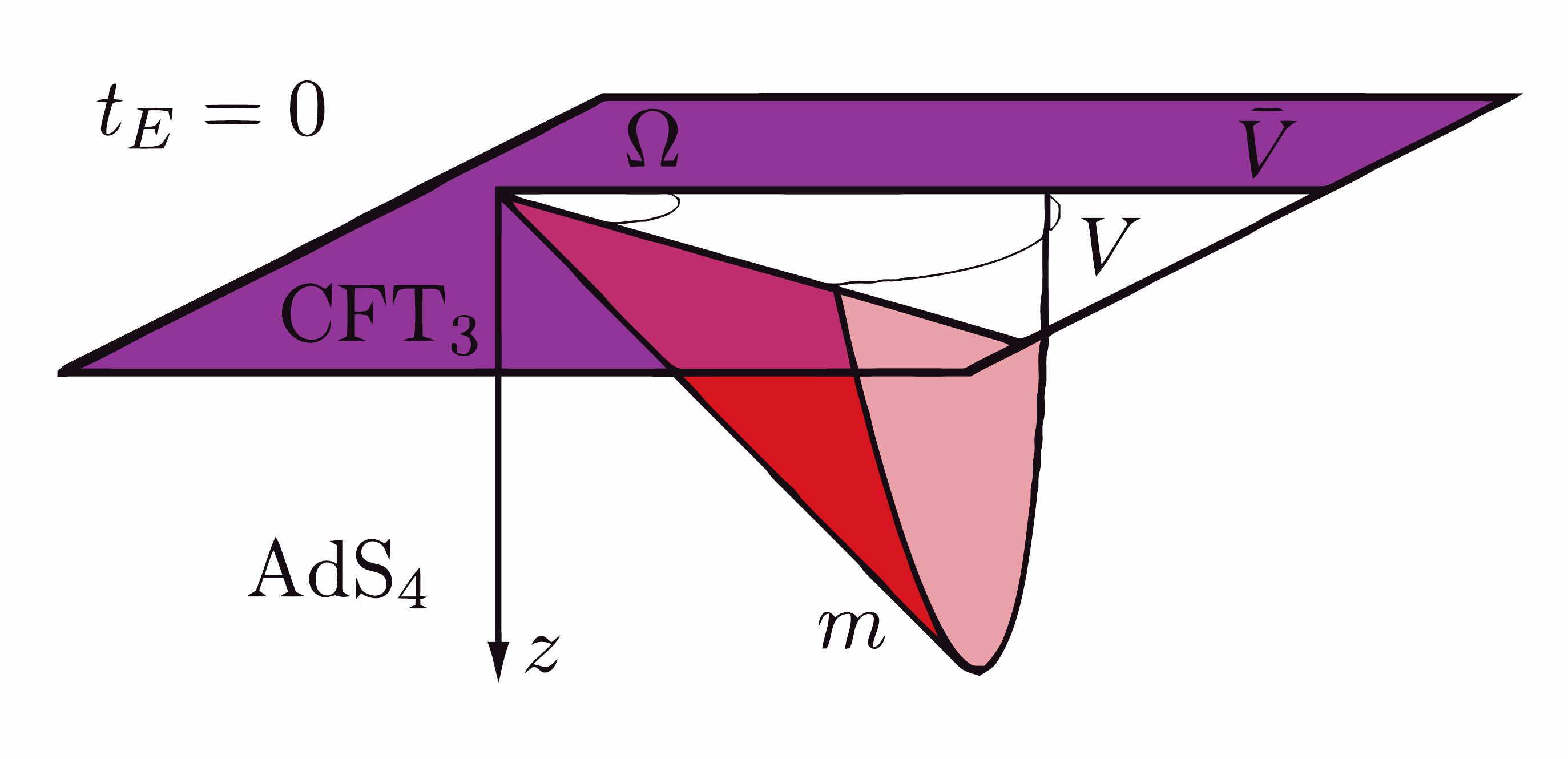}
\caption{(Colour online) A kink in a constant Euclidean time slice $t_E=0$ in the boundary of AdS$_4$.}
\labell{kinki}
\centering
\end{figure}
Now, the corresponding holographic entanglement entropy (HEE)  is computed using the Ryu-Takayanagi prescription for the entanglement entropy of conformal field theories dual to Einstein gravity \cite{Ryu:2006bv,Ryu:2006ef}.\footnote{This prescription has been recently proven under certain conditions in \cite{Lewk}.} According to this, the entanglement entropy of a certain region $V$ in our four-dimensional boundary theory is given by
\be 
\see(V)=\underset{m\sim V}{\text{ext}}\left[\frac{\mathcal{A}(m)}{4G} \right]\, ,\labell{RyuTaka}
\ee
where $m$ are codimension-$2$ bulk surfaces which are \emph{homologous} to $V$ in the boundary (and in particular $\partial m = \partial V$), and $\mathcal{A}(m)$ denotes the area  of $m$. Figure \ref{kinki} illustrates the extremal bulk surface for the region $V$ defined above.

Now following \cite{Hirata:2006jx}, we parametrize the bulk surfaces $m$ as $z=z(\rho,\theta)$ for the present case of corner region $V$. Further, the scaling symmetry of AdS, along with the fact that there is no other scale in the problem, allow us to limit the ansatz for the extremal surface to $z=\rho\, h(\theta)$, where $h(\theta)$ is a function satisfying $h\rightarrow 0$ as $\theta\rightarrow \pm \Omega/2 $. With this ansatz, the induced metric on the surface becomes
\begin{equation}
ds^2_{m}=\frac{\tilde{L}^2}{\rho^2}\left(1+\frac{1}{h^2} \right)d\rho^2+\frac{\tilde{L}^2}{h^2}\left(1+\dot{h}^2 \right)d\theta^2+\frac{2\tilde{L}^2\dot{h}}{\rho\, h}d\rho \,d\theta \, .\labell{indmee}
\end{equation}
where $\dot{h}\equiv dh/d\theta$. The entanglement entropy functional becomes then
\be
\see=\frac{1}{4G}\int_{m}d\theta\, d\rho\, \sqrt{\hm}=\frac{\tilde{L}^2}{2G}\int_{\delta/h_0}^{H}\frac{d\rho}{\rho}\,\int_{0}^{\Omega/2-\epsilon}\!\!\!d\theta\, \frac{\sqrt{1+h^2+\dot{h}^2}}{h^2} \, ,  \labell{eek}
\ee
where $\hm$ denotes the determinant of the induced metric (\ref{indmee}), we have introduced a UV cut-off at $z=\delta$ and $h_0\equiv h(0)$, which will be the maximum value of $h(\theta)$. As we already mentioned above, the $\rho$ integral is also cut-off as some large distance $H$. Finally, the angular cut-off $\epsilon$ is defined in such that at $z=\delta$, $\rho\, h(\Omega/2-\epsilon)=\delta$. Extremizing the above expression yields the equation of motion for $h(\theta)$, which reads
\be 
\ddot{h}(h+h^3)+h^4+3h^2+2(\dot{h}^2+1)=0\, . \labell{eomh}
\ee
However, the corresponding `Hamiltonian' is a conserved quantity, since there is no explicit $\theta$ dependence in eq.~\reef{eek}. Therefore we find the following first integral
\be 
\frac{1+h^2}{h^2\sqrt{1+h^2+\dot{h}^2}}=\frac{\sqrt{1+h_0^2}}{h_0^2} 
\, ,\labell{hhd}
\ee
where we used $\dot{h}(0)=0$. We can use eq.~(\ref{hhd}) to replace $\dot{h}$ in terms of $h$ and trade the integral over $\theta$ for one over $h$. After some algebra, eq.~(\ref{eek}) becomes
\begin{eqnarray} 
\see
 &=&\frac{\tilde{L}^2}{2G}\int_{\delta/h_0}^{H}\frac{d\rho}{\rho}\,\int_{0}^{\sqrt{(\rho/\delta)^2-1/h_0^2}}\!\!dy\ \sqrt{\frac{1+h_0^2(1+y^2)}{2+h_0^2(1+y^2)}} \, ,\labell{eek4}
\end{eqnarray}
where we have also substituted $y=\sqrt{1/h^2-1/h_0^2}$. Near the boundary ($y\rightarrow \infty$), the integrand behaves as
\be
 \sqrt{\frac{1+h_0^2(1+y^2)}{2+h_0^2(1+y^2)}}\sim 1 +\mathcal{O}\left(\frac{1}{y^2} \right)\, .
\ee
Therefore, the $y$ integration diverges in the limit that $\delta\to0$. However, we can isolate this divergence by adding and subtracting one to the integrand. Hence we recast eq.~(\ref{eek4}) as
\be 
\see=\frac{\tilde{L}^2}{2G}\int_{\delta/h_0}^{H}\frac{d\rho}{\rho}\,\int_{0}^{\infty}\!dy \left[\sqrt{\frac{1+h_0^2(1+y^2)}{2+h_0^2(1+y^2)}}-1 \right]+\frac{\tilde{L}^2}{2G}\int_{\delta/h_0}^{H}\frac{d\rho}{\rho} \sqrt{\frac{\rho^2}{\delta^2}-\frac{1}{h_0^2}}\, .\labell{eek3}
\ee
In the limit that $\delta\to0$, this expression can be further simplified to produce the final result
\be 
\see=\frac{\tilde{L}^2}{2G}\,\frac{H}{\delta}-\q(\Omega)\,\log\!\left(\!\frac{H}{\delta} \!\right)-\left(\frac{\pi\tilde{L}^2}{4G h_0}+\q(\Omega)\log (h_0)\right)+\mathcal{O}(\frac{\delta}{H}) \, ,\labell{eek5}
\ee
where the function $\q(\Omega)$ is given by
\be
\qe(\Omega)=\frac{\tilde{L}^2}{2G}\int_{0}^ {\infty} dy \left[1-\sqrt{\frac{1+h_0^2(1+y^2)}{2+h_0^2(1+y^2)}}\, \right]\, . \labell{q3}
\ee
The result in eq.~(\ref{eek5}) has precisely the expected form given in eq.~\reef{one}, \ie the first term in eq.~(\ref{eek5}) is, of course, the area law contribution, whereas the second is the universal contribution associated with the corner. The last one is the constant term, which does not have a universal character in the present situation. 

\begin{figure}[h]
        \centering
   		 \subfigure[ ]{
                \includegraphics[scale=0.56]{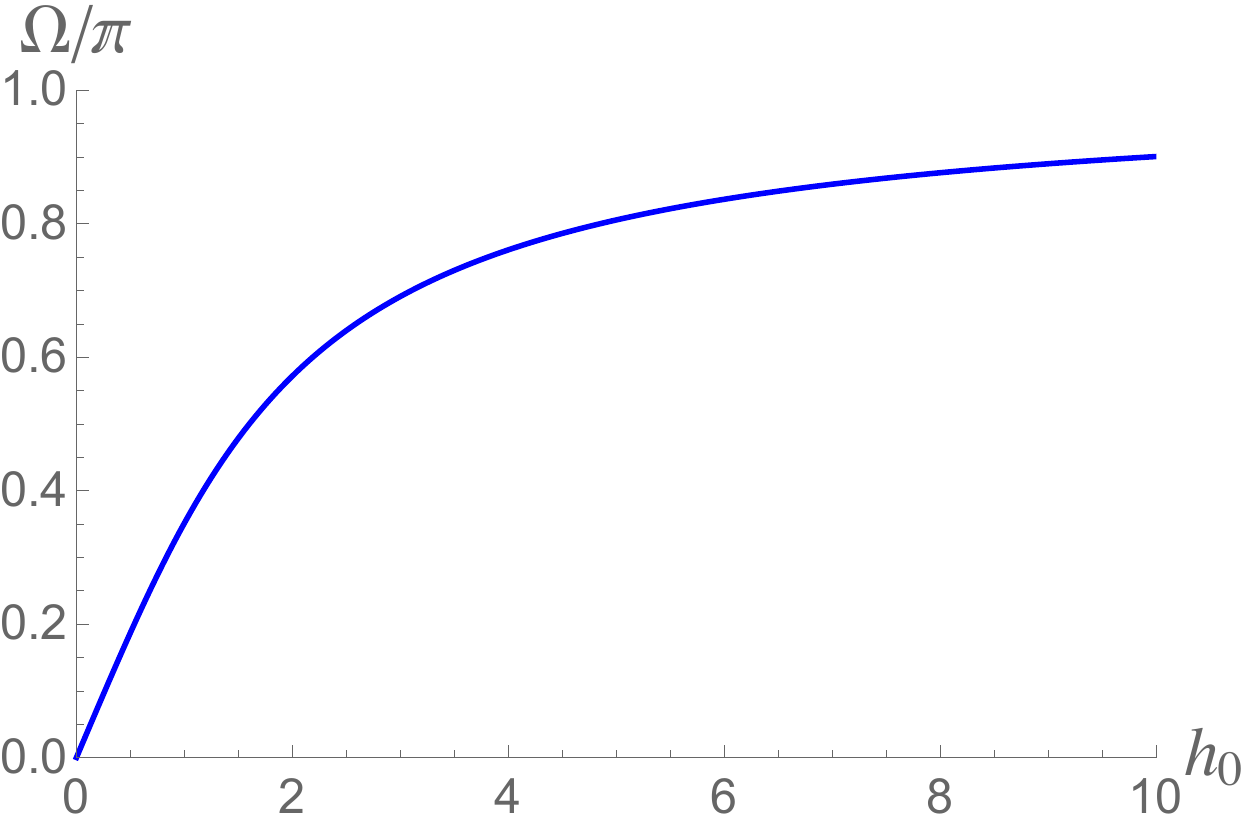}}
                \ \ \ 
         \subfigure[ ]{
                \includegraphics[scale=0.47]{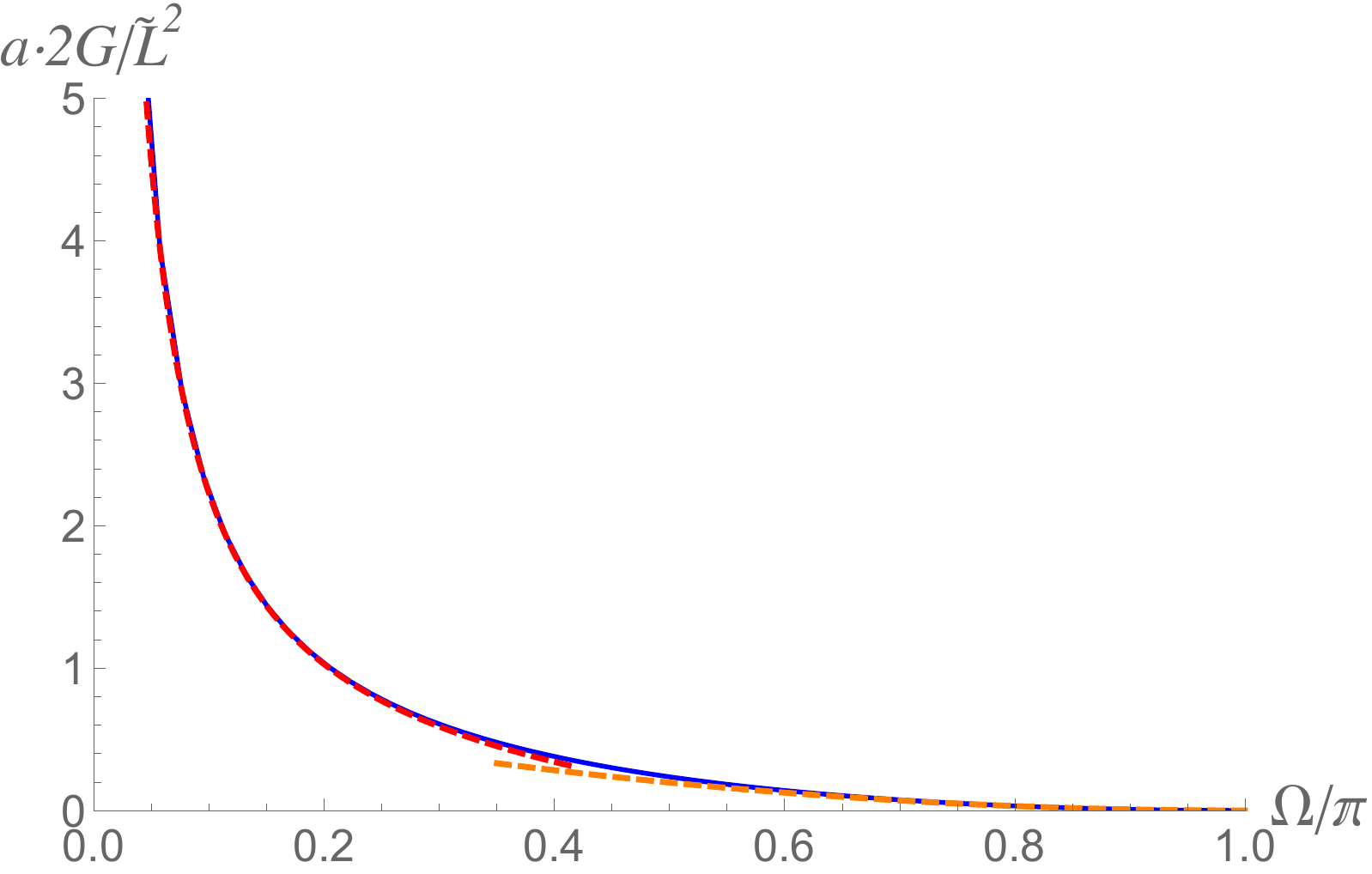}}
        \caption{(Colour online) (a) $\Omega/\pi$ as a function of $h_0$ and (b) ${\textstyle \frac{2G}{\tilde{L}^2}}\,\q$ as a function of $\Omega/\pi$. In the second panel, the dashed lines correspond to the approximate expressions derived in eqs.~\reef{lim1} and \reef{lim2} for small opening angles (red) and the smooth limit (orange), respectively.}
\labell{fig1}
\end{figure}
In eq.~\reef{q3}, we have added a subscript `$E$' to denote this function as the corner contribution with Einstein gravity in the bulk.
The dependence of $\qe(\Omega)$ on the opening angle is implicit on the right-hand side of eq.~\reef{q3} through the dependence of $h_0$ on $\Omega$. The latter can be determined by evaluating
\be \labell{om}
\Omega=\int_{-\Omega/2}^ {+\Omega/2}d\theta=\int_{0}^{h_0}dh\,\frac{2h^2\sqrt{1+h_0^2}}{\sqrt{1+h^2}\sqrt{(h_0^2-h^2)(h_0^2+(1+h_0^2)h^2)}}
\ee
and the result is shown in figure \ref{fig1}(a). The coefficient of the corner term is then plotted in figure \ref{fig1}(b) and we can see that $\qe(\Omega)$ does indeed satisfy all the various constraints explained in the introduction, \eg see eq.~\reef{restrict}. For small values of the opening angle, \ie $\Omega\rightarrow 0$, we find
\be
\Omega=\frac{2\sqrt{\pi}\, \Gamma\!\left(\frac{3}{4} \right) }{ \Gamma\!\left(\frac{1}{4} \right)}h_0-\frac{\left[3\Gamma\left(\frac{3}{4}\right)^2-\Gamma\left(\frac{1}{4}\right)\Gamma\left(\frac{5}{4}\right)\right]}{6\sqrt{2\pi}}h_0^3+\mathcal{O}(h_0^5)\, ,
\ee
\be \labell{lim1}
\qe(\Omega)=\frac{\tilde{L}^2}{2\pi G}\, \Gamma\!\left( {\textstyle \frac{3}{4}}\right)^4\,\frac{1}{\Omega}-\frac{\tilde{L}^2}{G}\frac{\pi\, \Gamma\!\left( \frac{1}{4}\right)}{48\sqrt{2}\,\Gamma\!\left( \frac{3}{4}\right)^3}\,\Omega+\mathcal{O}(\Omega^3)\, ,
\ee
which is shown as the dashed red line in figure \ref{fig1}(b)\footnote{Notice that eq. (\ref{lim1}) fits the exact $\qe(\Omega)$ curve remarkably well for not so small angles.}. Comparing the latter with eq.~\reef{ka}, we see that in this holographic model, the universal `central charge' associated with the small angle limit of the corner contribution is
\beq
{\rm Einstein\ gravity\,:}\quad \kae= \frac{\tilde{L}^2}{2\pi G}\, \Gamma\!\left( {\textstyle \frac{3}{4}}\right)^4\,.
\labell{gammaE}
\eeq
Considering the limit of a smooth entangling surface, \ie $\Omega\rightarrow \pi-\varepsilon$, we have
\be
\varepsilon=\frac{\pi}{h_0}+\mathcal{O}(h_0)\, ,
\ee
\be \labell{lim2}
\qe(\pi-\varepsilon)= \frac{\tilde{L}^2}{8\pi G}\,\varepsilon^2+\mathcal{O}(\varepsilon ^4)\, ,
\ee
which is shown as the dashed orange line in figure \ref{fig1}(b). Comparing this result with eq.~\reef{sigmA}, we see that the universal `central charge' associated with the limit of a nearly smooth entangling surface in this holographic model is
\beq
{\rm Einstein\ gravity\,:}\quad \sigma_{\ssc E}= \frac{\tilde{L}^2}{8\pi G}\,.
\labell{sigmaE} 
\eeq
Another interesting case to consider is a right-angled corner, \ie $\Omega=\pi/2$, for which we find
\be
\qe\left({\pi}/{2} \right)\simeq 0.11823\,\frac{\tilde{L}^2}{G}\simeq0.32944\,\kae\simeq 2.9714\,\sigma_{\ssc E} \,.\labell{right}
\ee
This case naturally arises in numerical calculations of entanglement entropy, \eg \cite{2014arXiv1401.3504K}.

\subsection{Higher curvature gravity}\labell{hcg}

Having reviewed the calculation for Einstein gravity in the bulk, we now turn to considering the effect of higher curvature interactions in the bulk theory. For such cases, the Ryu-Takayanagi prescription must be generalized, as was first considered in \cite{fur99,JHxb,highc2}. In particular, the Bekenstein-Hawking formula  on the right-hand side of eq.~\reef{RyuTaka} must be replaced by a new entropy functional which accounts for the new gravitational interactions. Hence eq.~\reef{RyuTaka} is replaced by
\beq
\see(V)=\underset{m\sim V}{\text{ext}}S_\mt{grav}(m)\, ,
\labell{newer}
\eeq
where the entropy functional $S_\mt{grav}$ depends on the details of the gravitational theory.
This is a familiar idea in the context of black hole entropy where the Wald entropy formula \cite{Wald:1993nt,ted9,wald2} extends $\mathcal{A}/(4G)$ with higher curvature corrections. A natural suggestion would be that the HEE should be calculated by extremizing the Wald entropy evaluated on the bulk surfaces $m$, however, it was shown that this approach would be incorrect since it fails to produce the proper universal contributions to the entanglement entropy \cite{JHxb}. The latter universal terms are properly reproduced in the special case of Lovelock gravity \cite{JHxb,highc2} using an alternative entropy functional \cite{Jacobson:1993xs}  --- see below. More generally the appropriate entropy functional is the Wald entropy plus additional terms involving the extrinsic curvature, which would vanish if evaluated on the Killing horizon of a stationary black hole \cite{RScalc,squash,Dong,Camps}. There has been an effort to extend the derivation \cite{Lewk} of the Ryu-Takayanagi prescription to higher curvature theories of gravity \cite{Dong,Camps,highc99,trouble} and a general formula was proposed for theories involving interactions with contractions of arbitrary powers of the Riemann tensor (but no derivatives of the curvature). While this general expression was shown to satisfy several consistency checks \cite{Dong}, it seems that it must still be further refined for general theories involving cubic and higher powers of the curvature \cite{trouble}. In any event, the correct entropy functional is known for general curvature-squared gravity in the bulk and we will use this to determine the modifications to the corner contribution in HEE for these theories in section \ref{cs}.

To go beyond curvature squared gravity, we turn to the generalized Lovelock theories considered by \cite{Sarkar}. In these theories, the Lagrangian is given by an arbitrary functional of extended `topological' densities, \ie scalars constructed from the curvature tensor which if integrated over a manifold of the appropriate dimension would yield the Euler characteristic. Hence Lovelock gravity \cite{Lovelock,lovel} would be the simplest example in which the Lagrangian is a linear functional of these topological densities. Another well-known class of theories which take this form would be $f(R)$ gravity \cite{revfr} since the Ricci scalar corresponds to the Euler density for two-dimensional manifolds. In studying these generalized Lovelock theories, \cite{Sarkar} proposed a formula for the gravitational entropy which satisfied a classical increase theorem for linearized perturbations of Killing horizons.\footnote{This result was recently extended to general higher curvature theories of gravity and a general connection was found with the entropy functional in HEE \cite{wall}.} We interpret the fact that their definition applies for at least small deviations away from a Killing horizon, as evidence that it will yield the correct gravitational entropy in the more general context of evaluating HEE. Then applying this prescription allows us to evaluate the modifications to the corner contribution in HEE for a certain class of theories involving cubic and higher powers of the curvature in section \ref{fl}.
 
Before proceeding with explicit calculations, let us comment that higher curvature interactions appear generically in string theoretic models, \eg as $\alpha^{\prime}$ corrections in the low-energy effective action \cite{gross}. However, rather than constructing explicit top-down holographic models, our approach here is to examine simple toy holographic models involving higher curvature interactions in the bulk gravity theory. Our perspective is that if there are interesting universal properties which hold for all CFTs, then they should also appear in the holographic CFTs defined by these toy models as well. This approach has been successfully applied before, \eg in the discovery of the F-theorem \cite{Myers:2010xs,Myers:2010tj}. We also stress that for the most part we will be working perturbatively in the gravitational couplings for the higher curvature interactions and only carry our calculations to first order in these couplings. The results for the curvature-squared theories are an exception, as most of these expressions are valid for generic values of the couplings.


\subsubsection{Curvature-squared gravity}\labell{cs}

The bulk action of the most general curvature-squared gravity can be written as
\begin{equation}
I_{2}=\frac{1}{16\pi G}\int d^4x\sqrt{g}\left[\frac{6}{L^2}+R+\lambda_1 L^2 R^2+\lambda_2 L^2 R_{\mu\nu}R^{\mu\nu}+\lgb L^2 \mathcal{X}_4 \right] \, ,\labell{actrr2}
\end{equation}
where
\begin{equation}
\mathcal{X}_4=R_{\mu\nu\rho\sigma}R^{\mu\nu\rho\sigma}-4R_{\mu\nu}R^{\mu\nu}+R^2 \, \labell{X4}
\end{equation}
is the Gauss-Bonnet term, \ie the Euler density for four-dimensional manifolds. Hence the last interaction does not effect the gravitational equations of motion since we are working with four bulk dimensions. However, as we will see, this term still contributes a topological term to the entropy functional. The AdS$_4$ metric in eq.~\reef{ads4} is still a solution of the full equations of motion for any value of $\lambda_1$ and $\lambda_2$ provided $\tilde{L}=L$.\footnote{This result is special to four dimensions. With a higher dimensional bulk, one would generally find $\tilde L^2=L^2/f_\infty$ where $f_\infty$ is a function of all three of the dimensionless couplings, $\lambda_1$, $\lambda_2$ and $\lgb$.}

The expression for the entanglement entropy in this family of theories is given by eq.~\reef{newer} where $S_\mt{grav}$ takes the form \cite{JHxb,RScalc,squash,Dong} 
\beq
\labell{see3} 
S_2=\frac{\mathcal{A}(m)}{4G} +\frac{L^2}{4G} \int_{m}\!\! d^2y\, \sqrt{\hm}\left[2\lambda_1  R+\lambda_2   \left(R^{\hat{a}}{}_{\hat{a}}-\frac{1}{2}K^{\hat{a}}K_{\hat{a}}\right) + 2 \lgb \mathcal{R} \, \right]\, ,
\eeq
where $\hm_{ij}$, $K^{\hat{a}}_{ij}$ and $\mathcal{R}$ are, respectively, the induced metric, the second fundamental form and the intrinsic Ricci scalar of the bulk surface $m$ --- see appendix \ref{app1} for a complete description of our conventions.\footnote{The last term in eq.~(\ref{see3}) corresponds to a particular case of the Jacobson-Myers entropy functional for Lovelock gravities \cite{Jacobson:1993xs}.} Before proceeding with detailed calculations of HEE, let us make some general observations about the expected results. 

First, it is worthwhile to note that the gravitational action \reef{actrr2} would also include various boundary terms, \eg see \cite{surf88,boundct}, and that similar boundary terms should be expected to appear in the entropy functional \reef{see3}. However, while the addition of such boundary terms may effect the coefficient in the area law contribution to the entanglement entropy \reef{one} in the boundary theory, one can infer from the local geometric form of these boundary terms that they will not modify the logarithmic contribution to $\see$ \cite{JHxb}. Again, the robustness of the logarithmic term here is a reflection of the fact that it is a universal contribution whose value is independent of the precise details of the UV regulator. Of course, since our interest lies in determining the universal corner term $\q(\Omega)$, we will ignore any boundary terms that might be added to eq.~\reef{see3}.

Next, let us examine the form of the entropy functional in eq.~\reef{see3}. The $\lambda_1$ and $\lambda_2$ terms both contain contributions involving the curvature of the background spacetime geometry. However, since we are evaluating the HEE in empty AdS$_4$, the latter terms are just constants, \eg $R=-12/{\tilde{L}}^2$. Hence the entropy functional is not modified by these terms except for a shift in the overall factor multiplying the area of bulk surface.\footnote{This simple shift may not arise when we are evaluating HEE in more general backgrounds, but this is a general result for backgrounds which are Einstein geometries, \ie $R_{\mu\nu}=-3/{\tilde L}^2\,g_{\mu\nu}$.} 

We also note that any surface which extremizes the area, as in eq.~\reef{RyuTaka}, will satisfy $K^{\hat{a}}=\hm{}^{ij}K^{\hat{a}}_{ij}=0$. Now looking at eq.~\reef{see3}, we see that the $\lambda_2$ contribution includes a term that is quadratic in $K^{\hat{a}}$. Hence an extremal area surface will also be a saddle point of this term. That is, if we deform away from the extremal area surface by some deformation parameterized by a small parameter $\varepsilon$, then we will have $K^{\hat{a}} \sim \mathcal{O}(\varepsilon)$ and $K^{\hat{a}}K_{\hat{a}}\sim \mathcal{O}(\varepsilon^2)$. Therefore extremal area surfaces will also extremize the new contribution (or any other contribution) to the HEE functional that is quadratic in the trace of the extrinsic curvature.\footnote{The full equations arising from extremizing the new functional will be very non-linear in general and so there may be other saddle points for which $ K^{\hat{a}}\neq 0$. However, we will also demand that the bulk surfaces reduce to the corresponding extremal area surfaces in the limit that $\lambda_i\rightarrow 0$. Therefore, these new non-linear solutions (if they exist at all) would be discarded since they would not satisfy this condition.} 

Lastly since we are working with a four-dimensional bulk spacetime, $m$ will be a two-dimensional manifold and hence $\int_m\!\! \sqrt{\hm}\, \mathcal{R}$, appearing as $\lgb$ contribution in eq.~\reef{see3}, will be proportional to a topological invariant (namely, the Euler characteristic) of $m$, up to boundary terms. Therefore just as the corresponding interaction in the bulk action \reef{actrr2} does not modify the gravitational equations of motion, this term in the HEE functional  will not contribute to the equations determining the bulk surface which extremizes eq.~\reef{see3}.  

Given the above discussion, we conclude that the extremal area surface for any given entangling region in the boundary of pure AdS$_4$ will also extremize the HEE functional \reef{see3} for the same calculation of entanglement entropy in the boundary theory dual to curvature-squared gravity. The only effect of the `higher curvature' corrections in eq.~\reef{see3} will be to change the final entanglement entropy by an overall factor depending on the new couplings, $\lambda_1$, $\lambda_2$ and $\lgb$. In the problem of interest, this indicates that the corner coefficient $\q(\Omega)$ will only be changed by this same overall factor.  Hence the expressions for the corner charges in eqs.~\reef{gammaE} and \reef{sigmaE} are also multiplied by an overall factor  but the functional dependence of $\q(\Omega)$ on the opening angle is precisely the same as compared to Einstein gravity. We note that the above observations actually have broader applicability and that this result will apply to a wide class of theories beyond the special case of curvature-squared gravity --- we return to a discussion of this point in section \ref{discuss}. Let us now turn to the detailed calculations to see how the different contributions in eq.~\reef{see3} affect the universal corner term in HEE. \\

\subsubsection*{$R^2$ gravity}
If we focus on the simplest case of $R^2$ gravity, \ie set $\lambda_2=0=\lgb$, the gravitational entropy functional reduces to
\beq
\labell{seer2}
 S_{2}=\frac{\mathcal{A}(m)}{4G} +\frac{L^2\lambda_1}{2G} \int_{m}\!\! d^2y\, \sqrt{\hm}\, R
 =\frac{\mathcal{A}(m)}{4G}\,(1-24\lambda_1)
\, ,
\eeq
where we substituted $R=-12/{L}^2$ to produce the last expression. Therefore, as discussed above, the corresponding corner coefficient is simply multiplied by an overall factor relative to the Einstein gravity\footnote{As we describe in appendix \ref{appf}, these results can be straightforwardly extended to the case of a general $f(R)$ gravity.}
\beq
\q(\Omega)=\left(1-24  \lambda_1 \right)\,\qe(\Omega)
\labell{newc2}
\eeq
and the corresponding small angle charge becomes
\be
\kappa=\left(1-24  \lambda_1 \right)\,\kae\, .
\labell{newk2}
\ee

\subsubsection*{$R_{\mu\nu}R^{\mu\nu}$  gravity}

 In the case of $R_{\mu\nu}R^{\mu\nu}$  gravity, the HEE functional becomes
\begin{eqnarray}
\labell{seeRic}
S_{2}&=&\frac{\mathcal{A}(m)}{4G} +\frac{L^2\lambda_2 }{4G} \int_{m}\!\! d^2y\, \sqrt{\hm}  \left(R^{\hat{a}}{}_{\hat{a}}-\frac{1}{2}K^{\hat{a}}K_{\hat{a}}\right) \\ 
\notag &=& \frac{\mathcal{A}(m)}{4G}\left(1-6 \lambda_2 \right) -\frac{\lambda_2}{8G} \int_{m} d^2y \sqrt{\hm}\, \frac{\left[2(1+\dot{h}^2)+3 h^2+h^4+(h+h^3) \ddot{h} \right]^2}{\left(1+h^2+\dot{h}^2\right)^3} \, .
\end{eqnarray}
where the expression in the second line was produced by first substituting $R^{\hat{a}}{}_{\hat{a}}=g^\perp{}^{\mu\nu}R_{\mu\nu}=-6/{L}^2$ and by evaluating 
$K^{\hat{a}}K_{\hat{a}}$ for the bulk surface defined by $z=\rho\,h(\theta)$ --- see appendix \ref{app1} for details.
Varying the above expression will produce a nonlinear differential equation for $h(\theta)$ which, because of the last term, involves third and fourth order derivatives, as well as first and second order derivatives. However, as we explained above, the solution should still be the same extremal area surface which we found with Einstein gravity. The latter occurs because the geometric form of the equation determining the extremal area surface is precisely $K^{\hat{a}}=0$. Indeed comparing with eq.~\reef{eomh}, we see that the factor in the numerator of the last term above is precisely the equation determining the profile $h(\theta)$ with Einstein gravity. Because this factor is squared, the profile satisfying eq.~\reef{eomh} will also satisfy the full equation of motion coming from eq.~\reef{seeRic} and further, in evaluating the HEE, the last term will not contribute because this factor simply vanishes. Hence the HEE and in particular, the corner coefficient, is determined by the Bekenstein-Hawking term, as with Einstein gravity but now multiplied by an additional factor. Therefore the charge defined by the corner term as in eq.~\reef{ka} becomes simply
\begin{eqnarray}
\labell{seeric2}
 \ka=\left(1-6\lambda_2\right) \,\kae\, .
\end{eqnarray}

\subsubsection*{Gauss-Bonnet gravity}\labell{gb}

For pure Gauss-Bonnet gravity, eq.~(\ref{see3}) reduces to
\begin{eqnarray}
\labell{seegb}
S_{2}=\frac{\mathcal{A}(m)}{4G} +\frac{L^2\lgb}{2G} \int_{m} d^2y\, \sqrt{\hm}\, \mathcal{R}\, .
\end{eqnarray}
Above, we argued that the second term would not affect the profile of the bulk surface nor contribute to the universal corner contribution.
With the bulk profile $z=\rho\,h(\theta)$, it is not difficult to show that the combination $\sqrt{\hm}\mathcal{R}$ can be written as a total derivative (see appendix \ref{app1} for details)
\be
\sqrt{\hm}\,\mathcal{R}=\frac{d}{d\theta}\left[\frac{2}{\rho}\,\frac{\dot{h}}{ h \sqrt{1+h^2+\dot{h}^2}} \right]\, . \labell{pop}
\ee
In fact, this is sufficient to conclude that the universal corner contribution will be identical to that in eq.~\reef{q3}, as expected.

However, let us examine the contribution of the Gauss-Bonnet term to the HEE in more detail.
Using eq.~\reef{pop}, this contribution can be written now as
\beq
\Delta S_{\text{GB}}=\frac{L^2\lgb}{2G} \int_{m} d^2y\, \sqrt{\hm}\, \mathcal{R}=-\frac{{L}^2\lgb}{G}\int_{\delta/h_0}^{H}\frac{d\rho}{\rho} \left[ \frac{\dot{h}}{h\sqrt{1+h^2+\dot{h}^2}} \right]_{\theta=0}^{\theta=\Omega/2-\epsilon}\, .
\labell{gbone}
\eeq
We can make use of eq. (\ref{hhd}) to replace $\dot{h}$ in terms of $h$. By doing so, and recalling that $h(\Omega/2-\epsilon)=\delta/\rho$ and $h(0)=h_0$, the above expression reduces to
\be
\Delta S_{\text{GB}}=-\frac{{L}^2\lgb}{G}\frac{H}{\delta}+\mathcal{O}(1)\, .
\ee
Hence, including the Einstein gravity, the final result for the HEE in this case becomes
\begin{eqnarray}
\labell{seeGB6}
\see=\frac{{L}^2}{2G}\frac{H}{\delta}\left(1-2\lgb \right)-\qe(\Omega)\, \log\! \left(\frac{H}{\delta} \right)+\mathcal{O}(1)  \, .
\end{eqnarray}
Hence the (nonuniversal) coefficient of the area law term has be modified here but the corner contribution is precisely the same as with just Einstein gravity in the bulk.

It was commented above that the entropy functional \reef{see3} might be supplemented by boundary terms but that the logarithmic term in the HEE, \ie the corner contribution,  is unaffected by such terms \cite{JHxb}. Gauss-Bonnet gravity provides an illustrative example since there is a natural boundary term to be added to the gravitational entropy functional  \cite{JHxb}
\begin{equation}
\labell{seeGB4}
S_{2}=\frac{\mathcal{A}(m)}{4G} +\frac{L^2\lgb}{2G} \int_{m} d^2y\, \sqrt{\hm}\, \mathcal{R}
+\frac{L^2\lgb}{G}\int_{\partial m}\!\! dy \sqrt{\tilde{\hm}}\, \mathcal{K}\, ,
\end{equation}
where $\partial m$ is the one-dimensional boundary of $m$ at the cut-off surface $z=\delta$. Further $\tilde{\hm}$ and $\mathcal{K}$ denote the determinant of the induced metric and the trace of the extrinsic curvature, respectively, on this boundary. It is straightforward to evaluate these quantities and to produce the result
\be \labell{om2}
\Delta S_{\text{GB}}^{\,'}=\frac{L^2\lgb}{G}\int_{\partial m}\!\! dy \sqrt{\tilde{\hm}}\, \mathcal{K}=\frac{L^2\lgb}{G}\int^{H}_{\delta/h_0} \frac{d\rho}{\delta}=\frac{L^2\lgb}{G}\,\frac{H}{\delta}+\mathcal{O}(1)\, .
\ee
Adding this contribution to eq.~(\ref{seeGB6}) leaves
\beq
\labell{seeGB7}
\see=\frac{{L}^2}{2G}\frac{H}{\delta}-\qe(\Omega)\, \log\! \left(\frac{H}{\delta} \right)+\mathcal{O}(1)  
\eeq
and we see that with the additional boundary term in eq.~\reef{seeGB4}, there is no $\lgb$ dependence in either the area law term or the logarithmic contribution in the entanglement entropy. The latter reflects the fact that with the additional boundary term, the Gauss-Bonnet contribution in eq.~\reef{seeGB4} is a purely topological contribution. In any event, as expected, the universal corner contribution remains unaffected by the addition of this boundary term, which implicitly represents a modification of the regulator used to define the entanglement entropy in the dual QFT.\\

To summarize our results for curvature-squared gravity \reef{actrr2} in the bulk, we found that the functional form of $\q(\Omega)$ is not modified. Rather the holographic expression only differs from that in the Einstein gravity by some overall factor. Hence the charge defined by the small $\Omega$ limit, as in eq.~\reef{ka}, becomes 
\begin{eqnarray}\labell{cccs}
\ka&=&\left(1-24\lambda_1-6\lambda_2 \right)\,\kae\,,
\end{eqnarray}
where the Einstein charge $\kae$ is given eq.~\reef{gammaE}.

\subsubsection{Generalized Lovelock gravity}\labell{fl}

Recall that Lovelock gravities \cite{Lovelock,lovel} are the most general higher curvature gravity theories with second-order equations of motion. The corresponding action can be written as 
\be
I_{\text{LL}}=\frac{1}{16\pi G}\int d^{d+1}x \, \sqrt{g} \,\left[\frac{d(d-1)}{L^2}+R+ \sum_{p=2}^{\left\lfloor\frac{d+1}{2}  \right\rfloor} \lambda_p L^{2p-2}\mathcal{L}_{2p}(R)\right]\, , \labell{LLact}
\ee
where $\lambda_p$ are dimensionless couplings and $\mathcal{L}_{2p}$ correspond to the dimensionally extended $2p$-dimensional Euler densities
\be
\mathcal{L}_{2p}(R)\equiv \frac{1}{2^p}\delta^{{\nu_1}{\nu_2}...{\nu_{2p-1}}{\nu_{2p}}}_{{\mu_1}{\mu_2}...{\mu_{2p-1}}{\mu_{2p}}} \,R^{{\mu_1}{\mu_2}}\,_{{\nu_1}{\nu_2}}\cdots R^{{\mu_{2p-1}}{\mu_{2p-2}}}\,_{{\nu_{2p-1}}{\nu_{2p-2}}}\, .
\ee
Here $\delta^{{\nu_1}{\nu_2}...{\nu_{2p-1}}{\nu_{2p}}}_{{\mu_1}{\mu_2}...{\mu_{2p-1}}{\mu_{2p}}}$ denotes a totally antisymmetric product of $2p$ Kronecker deltas. Hence when $p=(d+1)/2$, $\mathcal{L}_{2p}$ is topological and when $p>(d+1)/2$, $\mathcal{L}_{2p}$ simply vanishes. Of course, the cosmological constant and Einstein terms in eq.~\reef{LLact} could be incorporated into the sum as $\mathcal{L}_0$ and $\mathcal{L}_2$, respectively. Recently, there has been renewed interest in these theories in the context of the AdS/CFT correspondence where these theories provide toy models of holographic CFT's in which the central charges differ from one another, \eg see \cite{zoom} and the references therein. For this class of theories \reef{LLact},  HEE is evaluated with eq.~\reef{newer} using the following entropy functional \cite{JHxb,highc2}
\be
S_{\text{JM}}=\frac{\mathcal{A}(m)}{4G} + \frac{1}{4G}\int_m d^{d-1}y \, \sqrt{\hm}\sum_{p=2}^{\left\lfloor\frac{d+1}{2}  \right\rfloor} p\, \lambda_p\, L^{2p-2}\mathcal{L}_{2p-2}(\mathcal{R})\, , \labell{jm}
\ee
where now $\mathcal{L}_{2p-2}(\mathcal{R})$ is constructed with the intrinsic curvature tensor of the induced metric on $m$. 

Recently, Sarkar and Wall proposed a generalization of the Lovelock theories with an action of the form \cite{Sarkar} 
\be
I_{\text{SW}}=\frac{1}{16\pi G}\int d^{d+1}x \, \sqrt{g} \, f(\mathcal{L}_{0},\mathcal{L}_{2},\mathcal{L}_{4},\cdots,\mathcal{L}_{2k})\,, \labell{SWact}
\ee
where $f$ is some general function of the extended Euler densities up to $k=\lfloor (d+1)/2\rfloor$ --- we will assume that $f$ is a polynomial. Hence these new generalized Lovelock theories might also be seen as an extension of $f(R)$ gravity \cite{revfr}. In general, the gravitational equations of motion will involve fourth order derivatives of the metric in these new theories. However, the motivation to considering these theories is to examine the second law of black hole thermodynamics in higher curvature theories. In fact, \cite{Sarkar} found an expression for the gravitational entropy which satisfies a classical increase theorem for linearized perturbations of Killing horizons
\be\labell{sw}
S_{\text{SW}}=\frac{1}{4G}\int d^{d-1}y \, \sqrt{\hm}\sum_{p=1}^{\left\lfloor\frac{d+1}{2}  \right\rfloor} p\,\frac{\partial f}{\partial \mathcal{L}_{2p}(R)}\, \mathcal{L}_{2p-2}(\mathcal{R}_m)\, .
\ee
Certainly, this expression also reduces to that in eq.~\reef{jm} when $f$ is linear and the action \reef{SWact} is simply the Lovelock action \reef{LLact}.
We take these facts, in particular, that eq.~\reef{sw} applies for (at least small) deviations away from a Killing horizon, as evidence that it will yield the correct gravitational entropy in the more general context of using eq.~\reef{newer} to evaluate HEE. Further work in this direction recently appeared in \cite{wall}.

Hence we will use the generalized Lovelock theories \reef{SWact} as framework to examine the corner contribution in HEE. Since we are working in a four-dimensional bulk spacetime, all of the $\mathcal{L}_{2p}$ with $p=3,4,...$ will vanish identically. Therefore, we can only construct the new gravity action with powers of the Ricci scalar $R$ and the four-dimensional Euler density $\mathcal{X}_4$, given in eq.~\reef{X4}. Hence we consider supplementing the standard cosmological constant and Einstein terms in eq.~\reef{Einact} with higher curvature interactions of the form
\be\labell{actionlove}
\triangle I_{v,w}=\frac{\lambda_{v,w}}{16 \pi G}\int d^{4}x\,\sqrt{g}\, L^{2v+4w-2}\,R^v \,\mathcal{X}_4^w\, ,
\ee
with integers $v,w\geq1$. Then using eq.~(\ref{sw}), the corresponding entropy functional becomes
\be\labell{triage0}
\triangle S_{v,w}=\frac{\lambda_{v,w} }{4 G}\int_m \!\!d^{2}y\,\sqrt{\hm}\, L^{2v+4w-2}\,\left[v\, R^{v-1}\mathcal{X}_4^w+2w\,R^v\mathcal{X}_4^{w-1}\, \mathcal{R} \right]\, .
\ee
Now we are evaluating this expression in a pure AdS$_4$ background \reef{ads4} and so it may be simplified by substituting $R=-12/\tilde{L}^2$ and $\mathcal{X}_4=24/\tilde{L}^4$ to yield
\be\labell{triage}
\triangle S_{v,w}=(-1)^{v-1}\,2^{2v+3w-4}\, 3^{v+w-1}\,\frac{\lambda_{v,w} }{ G}\,\int_m d^{2}y\sqrt{\hm}\, \left[v-w L^2\,\mathcal{R}\right]\,f_{\infty}^{\,v+2w-1}\, .
\ee
Note the power of $f_\infty=L^2/\tilde{L}^2$ appearing in the integrand above. We have kept this factor here to indicate that in general after solving the gravitational equations, one finds that the curvature scale $\tilde{L}$ no longer coincides with the scale $L$ set by the cosmological constant. In particular, we find
\be \labell{llt}
1-f_{\infty}+(-1)^{v}2^{2v+3w-2}3^{v+w-1}\left(2-v-2w\right)\lambda_{v,w}\,f_{\infty}^{v+2w}=0\, .
\ee
However, note that if we are working perturbatively in the coupling, we have 
\be \labell{lltt}
L^2=\tL^2 f_\infty\simeq \tL^2 \left[1+(-1)^{v}2^{2v+3w-2}3^{v+w-1}\left(2-v-2w\right)\lambda_{v,w}+{\cal O}(\lambda^2_{v,w})\right]\,.
\ee

With the simplifications produced by working in AdS$_4$, the modifications to the entropy functional have reduced to a term proportional to the area of the bulk surface and another involving an integral of the intrinsic Ricci scalar over $m$. Hence at this point, we can turn to our results from the previous subsection where both terms were encountered before. In particular, neither term modifies the profile of the extremal surface in the bulk and further the area term only changes the corner contribution by an overall factor while the term involving $\mathcal R$ does not contribute to this universal term at all. More precisely, given the precise results in eq.~\reef{triage}, we find that the small angle charge associated with the corner term becomes
\beq
\labell{chargeVW}
\ka=\left[1-(-1)^{v}\,2^{2v+3w-2}\, 3^{v+w-1}\,v\,\lambda_{v,w}+\mathcal{O}(\lambda^2_{v,w}) \right]\kae\,,\, \quad \kae=\frac{\tL^2}{2\pi G}\Gamma\left(\textstyle \frac{3}{4} \right)^4\, ,
\eeq
where the result is expressed to leading order in the perturbative expansion in the coupling. Note that we have 
expressed $\kae$ in terms of the AdS scale $\tL^2$, which differs here from the scale $L^2$ in the action by terms of $\mathcal{O}(\lambda_{v,w})$, as shown in eq.~\reef{lltt}. If we expressed the above equation in terms of $L^2$ instead, the $\mathcal{O}(\lambda_{v,w})$ coefficient would change. However, our convention here and throughout the following will be to write all of our perturbative expressions in terms of $\tL^2$. Of course, all length scales will disappear from the ratios of the different charges and so once our results are expressed in this way, they will not depend on this convention. Further, having fixed our approach, the comparison with the calculations for Einstein gravity is unambiguous in all cases. \\

To make our analysis more concrete, let us extend the general curvature-squared theory \reef{actrr2} with the generalized Lovelock interactions which are third- and fourth-order in the curvature 
\begin{eqnarray}\labell{fo}
&&I=\frac{1}{16\pi G}\int d^{4}x \, \sqrt{g}\left[\frac{6}{L^2}+R+L^2\left(\lambda_1 R^2+\lambda_2 R_{\mu\nu}R^{\mu\nu}+\lambda_{\text{GB}} \mathcal{X}_4\right)\right.\\ \notag 
&&\qquad\qquad+L^4 \left(\lambda_{3,0}R^3+\lambda_{1,1}R\mathcal{X}_4\right) +L^6 \left(\lambda_{4,0}R^4+\lambda_{2,1}R^2\mathcal{X}_4+\lambda_{0,2}\mathcal{X}_4^2\right) \bigg] \,\, .
\end{eqnarray}
Then the final expression of the corner coefficient and the corresponding charge take the simple form
\beq
\q(\Omega)=\alpha\,\qe(\Omega)\quad{\rm and}\quad \ka=\alpha\,\kae
\labell{finfin}
\eeq
where to leading order in the dimensionless couplings, the overall coefficient is given by
\beq
\labell{chacha}
\alpha=1-24\lambda_1-6\lambda_2+432\lambda_{3,0}+24\lambda_{1,1}-6912\lambda_{4,0}-576\lambda_{2,1}+\mathcal{O}(\lambda^2)\, .
\eeq
Of course, $\qe(\Omega)$ and $\kae$ are the corresponding quantities evaluated for Einstein gravity, as given in eqs.~\reef{q3} and \reef{gammaE}, respectively.
The fact that the functional form of $\q(\Omega)$ is unchanged results because the higher curvature contributions to the entropy functional studied here do not modify the profile of the extremal surface in the bulk. We do not expect that this behaviour is completely universal and it may be modified in theories with even more general higher curvature interactions. We will come back to this point in the discussion section.

\section{Comparison with other charges}\labell{witncc}

By considering the limit of a small opening angle in eq.~\reef{ka}, we identified two `central charges' which appear in the entanglement entropy of regions where boundary has corners. When evaluated for holographic CFT's dual to Einstein gravity, the result \reef{gammaE} is proportional to the ratio $\tL^2/G\sim \tL^2/\ell_{\ssc Planck}^2$. The latter ratio is well known to be indicative of the number of degrees of freedom in the boundary theory. However, for Einstein gravity, the same ratio is ubiquitous for physical quantities involving a similar count of degrees of freedom, \eg the entropy density of a thermal bath. The pervasiveness of $\tL^2/G$ arises since this is the only dimensionless parameter that is intrinsic to the bulk theory with Einstein gravity. By considering higher curvature theories for the bulk gravity, as in the previous section, we are introducing more dimensionless couplings and we can begin to distinguish the various charges in the boundary theory, \eg see \cite{JHxb,gb2,Myers:2010jv}. Our objective here is to use our holographic results to determine if the corner charge $\ka$ should be considered a new and distinct charge or if it is proportional to charges already appearing in other physical quantities. In particular, in the following, we compare $\ka$ to the analogous charges appearing in: 1) the entanglement entropy of an infinite strip; 2) the entanglement entropy of a disk; 3) the entropy density of a thermal bath and 4) the two-point function of the stress tensor. Again, with Einstein gravity in the bulk, all of these quantities are proportional to $\tL^2/G$. While the same is true (with our conventions) with the higher curvature theories, the additional dimensionless couplings also give each a unique signature, as we will see in the following.

\subsection{Entanglement entropy for a strip}\labell{belt}

We begin with the entanglement entropy of an infinite strip. For a general three-dimensional CFT, the result will take the form \cite{Casini:2009sr,gb}
\be
\see = c_1 \, \frac{2H}{\delta} - \ta\, \frac{H}{\ell}+\cO(\delta)\, \labell{strip}
\ee
where $\ell$ is the width of the strip and $H$ is a long distance scale introduced to regulate the length of the strip, \ie the area of the entangling surface is $2H$. The universal coefficient $\ta$ can be isolated with
\be
\ta =\frac{\ell^2}{H}\,\frac{\partial \see}{\partial \ell}\,.
\labell{charstrip}
\ee
We will find that $\ta=\ka$ in our HEE calculations below. In fact, this result holds for general three-dimensional CFT's and has a simple explanation since there is a conformal transformation that (essentially) relates the corresponding entanglement geometries --- see appendix \ref{app4}.

Holographic calculations of the entanglement entropy of a strip were first carried out in \cite{Ryu:2006bv,Ryu:2006ef} with Einstein gravity in the bulk. To start, we write AdS$_4$ metric as
\begin{equation}\labell{ads42}
ds^2=\frac{\tL^2}{z^2}\left(dz^2+d\te^2 +dx_1^2+dx_2^2  \right)\, .
\end{equation}
Let us parameterize the strip in the boundary as the region B$=\left\{\te=0,\, x_1\in[-\ell/2,\ell/2] \right\}$. As noted above, we also introduce an IR regulator by, \eg making the $x_2$ direction periodic with period $\triangle x_2=H$ and with $H\gg\ell$. The translational symmetry along $x_2$ allows us to parametrize the entangling surface $m$ as $z=h(x_1)$, so the induced metric on the surface becomes
\begin{equation}\labell{indS}
ds^2_{m}=\frac{\tilde{L}^2}{h^2}\left(\left[1+\dot{h}^2 \right]dx_1^2+ dx_2^2\right)\, ,
\end{equation}
where $\dot{h}=\partial_{x_1}h$. Focusing on Einstein gravity \cite{Ryu:2006bv,Ryu:2006ef,Hirata:2006jx}, we look for surfaces $m$ extremizing the area functional, which in this case is given by
\begin{equation}
\labell{seeGBd2}
 S_{B}=\frac{\tilde{L}^2}{4G}H \int_{-\ell/2}^{\ell/2} dx_1\, \frac{1}{h^2}\sqrt{1+\dot{h}^2} \, .
\end{equation}
Since the integrand does not depend on $x_1$ explicitly, there is conserved first integral which can be used to write
\begin{equation}
\labell{fid}
\dot{h}=-\frac{\sqrt{z_*^4-h^4}}{h^2}  \, ,
\end{equation}
where $z_*$ is the maximal value of $z$ reached by the extremal surface. The latter can be identified in terms of $\ell$ through
\begin{equation}
\ell=2\int_0^{\ell/2}dx_1=2\int_0^{z_*}\frac{h^2\,dh}{\sqrt{z_*^4-h^4}}=\frac{\sqrt{2}}{\sqrt{\pi}}\,\Gamma\left({\textstyle \frac{3}{4}}\right)^2\,z_*  \, .
\end{equation}
The final result for the entanglement entropy with Einstein gravity in the bulk is
\begin{equation}
\labell{seeGBd334a}
 S_{B}=\frac{\tilde{L}^2}{2G}\, \frac{H}{\delta}- \frac{\tilde{L}^2}{2\pi G} \,\Gamma\left({\textstyle \frac{3}{4}}\right)^4\,\frac{H}{\ell}   \, ,
\end{equation}
Hence the corresponding universal coefficient is
\be
\labell{tae}
\ta_{\ssc E}=\frac{\tilde{L}^2}{2\pi G} \,\Gamma\left({\textstyle \frac{3}{4}}\right)^4\,,
\ee
which exhibits the expected factor of $\tL^2/G$, and further comparing with eq.~\reef{gammaE}, we see that $\ta_{\ssc E}=\kae$.

This calculation of HEE is easily extended to the higher curvature theories considered in section \ref{hcg}, taking into account the general remarks made there. We use the prescription \reef{newer} with the generalized entropy functionals for those theories given in eqs.~\reef{see3} and \reef{triage0}. However, as we found before, the terms involving the trace of the extrinsic curvature do not contribute, those with the intrinsic Ricci scalar only contribute boundary terms  and those involving bulk curvatures only modify the Einstein result by an overall factor. It is straightforward to verify these expectations with explicit calculations and the final result is
\begin{equation}
\labell{seeGBd334}
 S_{B}=\beta\frac{\tilde{L}^2}{2G}\, \frac{H}{\delta}- \alpha\,\frac{\tilde{L}^2}{2\pi G} \,\Gamma\left({\textstyle \frac{3}{4}}\right)^4\,\frac{H}{\ell}\, ,
\end{equation}
where $\alpha$ is precisely the same factor given in eq.~(\ref{chacha}). The coefficient $\beta$ appearing in the area law term is another function of the couplings $\lambda_i$, which is not needed here but does not coincide with $\alpha$ in general.\footnote{In fact, the same factor $\beta$ appears below in the HEE calculation for a disk.} Hence the final result for the universal coefficient is
\be\labell{newta}
\ta=\alpha\,\ta_{\ssc E} \,,
\ee
and so we find that $\ta=\ka$ in all of these examples. As noted above, this is in fact a general result for three-dimensional CFT's.

\subsection{Entanglement entropy for a disk} \labell{disk}

For a general three-dimensional CFT, the entanglement entropy of a disk will take the form \cite{Casini:2011kv,mutual}
\be
\see = c_1 \, \frac{2\pi R}{\delta} - 2\pi\, c_0+\cO(\delta)\, \labell{disk}
\ee
where $R$ is the radius of the disk. The universal coefficient $c_0$ can be isolated here by evaluating \cite{markm}
\be
c_0 =\frac{1}{2\pi}\left(R\,\frac{\partial \see}{\partial R}-\see\right)\,.
\labell{chardisk}
\ee
Of course, in this case, the universal constant $c_0$ plays the an important role as the central charge in the $F$-theorem, \ie it decreases monotonically in renormalization group flows \cite{Myers:2010xs,Myers:2010tj,Fthem1,Fthem2,proof}.

The HEE for a disk was first calculated for Einstein gravity using eq.~\reef{RyuTaka} in \cite{Ryu:2006bv,Ryu:2006ef}. However, this calculation was later extended to general higher curvature theories of gravity in the bulk \cite{Myers:2010tj,Casini:2011kv}. Making use of a conformal transformation in the boundary CFT, the problem of calculating the entanglement entropy for a disk can be mapped to the question of evaluating the thermal entropy of the CFT in a particular curved background. The latter can then be evaluated as the Wald entropy of the corresponding horizon in bulk spacetime with a general gravitational theory in the bulk. The horizon actually appears as an `observer' horizon upon transforming the bulk AdS geometry to AdS-Rindler coordinates and the extremal area surface in the standard calculation coincides with the bifurcation surface of this horizon, \eg see \cite{eeom}. 
 
Our calculations of HEE for the disk followed the prescription outlined in section \ref{hcg}, using eq.~\reef{newer} with the entropy functionals in eqs.~\reef{see3} and \reef{triage0}. Using the AdS$_4$ metric in eq.~\reef{ads4}, let us parameterize the disk in the boundary as the region  $D=\left\{\te=0,\, \rho\leq R \right\}$. We write the profile of the bulk surface $m$ as $z=h(\rho)$ with no dependence on $\theta$ because of the rotational symmetry of the disk. The induced metric on $m$ then becomes 
\begin{equation}
ds^2_{m}=\frac{\tilde{L}^2}{h^2}\left(\left[1+\dot{h}^2 \right]d\rho^2+\rho^2 d\theta^2\right)\, ,
\end{equation}
where $\dot{h}=\partial_\theta h$. The extremal area surface becomes the hemisphere  \cite{Ryu:2006bv,Ryu:2006ef}
\begin{equation}\labell{bulkD}
\rho^2+z^2=R^2  \quad {\rm with}\ z\geq 0\, .
\end{equation}
Now in general, the entropy functional for higher curvature theories can be written as the Wald entropy plus terms which are at least quadratic in the extrinsic curvature \cite{Dong,Camps}. However, one can readily verify that the extrinsic curvature of the above bulk surface \reef{bulkD} vanishes and hence any extrinsic curvature terms will vanish to first order if we make variations of this surface. Since the Wald entropy only involves bulk curvatures, this entropy reduces to the area functional multiplied by an extra overall factor, as in the previous section. Hence eq.~\reef{bulkD} still remains the extremal surface when calculating the HEE of a disk for any general higher curvature theory in the bulk. Hence with eqs.~\reef{see3} and \reef{triage0} for the theories in section \ref{hcg}, evaluating the HEE yields
\begin{equation}
\labell{seeR2ss}
 S_{D}=\beta\,\frac{\pi \tilde{L}^2}{2 G} \,\frac{R}{\delta}-\beta\,\frac{\pi\tilde{L}^2}{2 G}    \, ,
\end{equation}
where \footnote{For a general theory with action (\ref{actionlove}), the corresponding expression is $$\beta=1+(v+w)(-1)^{v-1}2^{2v+3w-2}3^{v+w-1}\lambda_{v,w}+\mathcal{O}(\lambda_{v,w}^2)\,.$$}
\begin{equation}
\beta=1-24\lambda_1-6\lambda_2-2\lgb+432\lambda_{3,0}+48\lambda_{1,1}-6912\lambda_{4,0}-864\lambda_{2,1}-96\lambda_{0,2}+\mathcal{O}(\lambda^2)\, .\labell{betaD}
\end{equation}
Hence the universal charge for the corresponding holographic CFT's becomes
\be
\labell{charDD}
c_0= \beta \,c_{0,{\ssc E}}=\beta\,\frac{\tilde{L}^2}{4 G}\,,
\ee
where $c_{0,{\ssc E}}$ denotes the result for Einstein gravity, \ie $c_{0,{\ssc E}}=\tL^2/(4 G)$. Note that with Einstein gravity, the ratio of the universal charges for the corner and the disk is relatively simple, \ie
\be
\frac{\kae}{c_{0,{\ssc E}}}=\frac{2}{\pi}\Gamma\!\left({\textstyle \frac{3}{4}}\right)^4\,. \labell{ratioE1}
\ee
However, comparing eqs.~\reef{newta} and \reef{charDD}, as well as eqs.~\reef{chacha} and \reef{betaD}, we see that there is no simple relation between $\ka$ and $c_0$ in the general theories. In particular, we have
\be
\frac{\ka}{c_{0}}=\frac{2}{\pi}\,\Gamma\!\left({\textstyle \frac{3}{4}}\right)^4\,\left(1-2\lgb-24\lambda_{1,1}+288\lambda_{2,1}+96\lambda_{0,2}+\mathcal{O}(\lambda^2) \right) \labell{ratiokc}
\ee
and so this ratio depends on the precise value of the gravitational couplings in the higher curvature theories.

\subsection{Thermal entropy}\labell{thermal}

Another quantity which might be used to characterize the number of degrees of freedom in a system is the thermal entropy. For a three-dimensional CFT, the thermal entropy density takes the form
\be
s= 
c_{\ssc S}\,T^2\,.\labell{thermS}
\ee
The coefficient $\cs$ is another interesting `central charge' which is readily calculable in a holographic setting. Of course, the thermal bath in the boundary theory is dual to a planar AdS$_4$ black hole and we need only calculate the entropy density of the event horizon. For Einstein gravity, the black hole solution can be written as 
\begin{equation}\labell{bb}
ds^2=\frac{\tilde{L}^2}{z^2}\left(\frac{dz^2}{f(z)}-f(z) dt^2+dx_1^2+dx_2^2 \right) \quad{\rm with}\ 
f(z)\equiv1-\frac{z^3}{z_\text{H}^3}\, ,
\end{equation}
where $z=z_\text{H}$ is the position of the event horizon. The Hawking temperature is given by $T=3/(4\pi z_\text{H})$ and  the horizon entropy is given by the Bekenstein-Hawking formula, which yields
\begin{equation}\labell{tebb}
S_{\text{thermal}}=\frac{1}{4G} \int_{z=z_\text{H}} \sqrt{h}\, d^2{x}=\frac{\tL^2}{4G\,z_\text{H}^2}\,V_2  \, ,
\end{equation}
where $V_2\equiv \int dx_1 dx_2$. Now dividing by the spatial volume $V_2$ yields the entropy density and substituting the temperature for $z_\text{H}$ produces an expression of the expected form given in eq.~\reef{thermS}. 
The corresponding central charge is
\be
\cse=\frac{4\pi^2}{9}\,\frac{\tilde{L}^2}{G}\,.
\labell{chargeSE}
\ee
Here again, we see the ubiquitous factor of $\tL^2/G$ and hence the ratio with the corner charge yields a fixed numerical factor, \ie
\be
\frac{\kae}{\cse}=\frac{9}{8\pi^3}\,\Gamma\!\left({\textstyle \frac{3}{4}}\right)^4\,. 
\labell{ratioE2}
\ee

\subsubsection*{Curvature-squared gravity}

Just as with empty AdS$_4$, the black hole metric \reef{bb} is also a solution of the general curvature-squared gravity for any value of the parameters $\lambda_1$, $\lambda_2$ and $\lambda_{\text{GB}}$ provided $\tilde{L}^2=L^2$. Hence the only difference from the above calculations is that the horizon entropy is now given by the Wald entropy formula \cite{Wald:1993nt,ted9,wald2}. Alternatively, we can use the generalized entropy functional in eq.~\reef{see3} since the two expressions only differ by terms involving the extrinsic curvature and the latter vanishes on the event horizon of the AdS$_4$ black hole. We find, in agreement with \cite{Smolic:2013gz}
\begin{equation} 
s=(1-24\lambda_1-6\lambda_2)\, \frac{4\pi^2 \tilde{L}^2}{9\,G}\,T^2\labell{sqs1}
\end{equation}
and therefore the corresponding central charge becomes
\be
\cs=\gamma_2\, \cse\qquad{\rm with}
\ \ \gamma_2=1-24\lambda_1-6\lambda_2\,.
\labell{sqs2}
\ee
Comparing to eq.~\reef{cccs}, we see that for curvature-squared gravity, the thermal entropy charge is modified by the same overall factor that appears in the corresponding corner charge. Hence for this family of holographic theories, the ratio of these two charges remains unchanged from the numerical factor \reef{ratioE2} that appears with Einstein gravity.  

\subsubsection*{Generalized Lovelock gravity}

The black hole metric in eq.~(\ref{bb}) is no longer a solution of the equations of motion for general theories of the form (\ref{actionlove}). Hence in order to explore how the thermal entropy gets modified here, we must first correct the black hole solution to linear order in the coupling $\lambda_{v,w}$. We parametrize the modified solution as
\begin{equation}\labell{bb6}
ds^2=\frac{\tilde{L}^2}{z^2}\left(\frac{dz^2}{f(z)\left[1+\lambda_{v,w} f_2(z) \right]}-f(z)\left[1+\lambda_{v,w} f_1(z) \right] dt^2+dx_1^2+dx_2^2 \right)\, ,
\end{equation}
where $f_1(z)$ and $f_2(z)$ are two nonsingular functions to be determined. This ansatz was chosen so that the position of the horizon remains at $z=z_{\text{H}}$. In order to obtain $f_1(z)$ and $f_2(z)$, we substitute the above metric into the Einstein action (\ref{Einact}) modified by the addition of a higher curvature interaction as in eq.~\reef{actionlove} and expand to second order in the coupling $\lambda_{v,w}$.\footnote{Since we are working perturbatively in $\lambda_{v,w}$, it is sufficient to consider each higher curvature interaction \reef{actionlove} separately. Of course, the first order variations by $f_1(z)$ and $f_2(z)$ vanish identically here because to leading order, the metric solves the Einstein equations of motion.} From the second order action, we determine the linearized equations of motion for $f_1(z)$ and $f_2(z)$ and then solve them with the boundary conditions that both functions decay as $z\to0$ and remain nonsingular at $z=z_{\text{H}}$. Below we describe the solution and the results for the thermal entropy for each of the generalized Lovelock interactions up to quartic order in the curvatures, shown in eq.~\reef{fo}.

In general, the Hawking temperature of the solution will be given by
\begin{equation}
T=\frac{3}{4\pi z_\text{H}}\left(1 + \frac{f_1(z_\text{H})+f_2(z_{\text{H}})}{2}\lambda_{v,w}+\mathcal{O}(\lambda_{v,w}^2)\right) \, ,
\end{equation}
as one can easily check.

\subsubsection*{a) $R^3$ and $R^4$ gravity}

For these two particular theories, as well as any theory with only $R^v$ interactions (\ie $w=0$), the original AdS$_4$ black hole solution \reef{bb} does not get corrected at any order in the couplings $\lambda_{v,0}$, \ie $f_1(z)=f_2(z)=0$. The uncorrected black hole solves the equations of motion of these theories provided the curvature scale satisfies eq.~(\ref{llt}), which was also required for the pure AdS$_4$ metric \reef{ads4} to be a solution in the new theory. Note that for $v=3$ and 4, we find the constraints $1-f_{\infty}+144\lambda_{3,0}f_{\infty}^3=0$  and $1-f_{\infty}-3456\lambda_{4,0}f_{\infty}^4=0$, respectively.

The horizon entropy is computed using the expression in eq.~(\ref{sw}). However, since the Ricci scalar of the Schwarzschild-AdS$_4$ background equals that of the pure AdS$_4$ solution, the corrected thermal entropy for these theories differs from the Einstein gravity result by just a overall constant factor which is precisely the same as the $\lambda_{3,0}$ and $\lambda_{4,0}$ contributions to $\alpha$ in eq.~(\ref{chacha}). That is, we find
\be\labell{grama}
s=\gamma_{a}\,\cse\,T^2 \qquad{\rm with}\ \ \gamma_{a}= 1+432\lambda_{3,0}-6912\lambda_{4,0}+\mathcal{O}(\lambda^2) \, .
\ee

\subsubsection*{b) $R\,\mathcal{X}_4$ gravity}

For this theory, the AdS curvature is given by $1-f_{\infty}+24\lambda_{1,1} f_{\infty}^3=0$ --- recall that $\fin\equiv L^2/\tL^2$.
The planar black hole \reef{bb} no longer solves the equations of motion and so we proceed as described above to find the corrected solution to first order in the coupling. The two functions $f_1$ and $f_2$ are 
\begin{eqnarray}\labell{f1f2}
f_1(z)&=&-\frac{18z^3(z^3+z_\text{H}^3)}{z_\text{H}^6}\, ,\\ \notag
f_2(z)&=&\frac{6z^3(11z^3-3z_\text{H}^3)}{z_\text{H}^6}\, .
\end{eqnarray}
With the new metric, the Hawking temperature becomes
\be \labell{T6}
T_{1,1}=\frac{3}{4\pi z_\text{H}}\left(1+6\lambda_{1,1}+\mathcal{O}(\lambda_{1,1}^2)\right)\, .
\ee
 Using eq.~\reef{triage0}, the thermal entropy then becomes
\be\labell{gramb}
s=\gamma_{b}\,\cse\,T^2 \qquad{\rm with}\ \ \gamma_{b}= 1+24\lambda_{1,1}+\mathcal{O}(\lambda_{1,1}^2)\, .
\ee
We note that $\gamma_b$ again agrees with the analogous factor appearing in the corner coefficient \reef{chargeVW} for $v=1=w$. 

We stress that, as opposed to the theories with $w=0$, the on-shell Gauss-Bonnet term $\mathcal{X}_4$ is no longer the same in the black hole background as in the pure AdS$_4$ solution (hence eq.~\reef{triage0} no longer reduces down to eq.~\reef{triage}). Computing the horizon entropy as a function of the horizon position yields
\begin{equation}
s=(1+36\lambda_{1,1}+\mathcal{O}(\lambda_{1,1}^2))\frac{ \tilde{L}^2}{4\pi G z_\text{H}^2}\, .
\end{equation}
It is only when we express the entropy density as a function of the physical temperature \reef{T6} that we cover the factor $\gamma_{b}$ in eq.~\reef{gramb}. Actually, it is possible to show that different parametrizations of the corrected solution give rise to different expressions for $s(z_\text{H})$ and $T(z_\text{H})$, which nevertheless conspire to produce the same physical result when the entropy density is written in terms of the temperature.

\subsubsection*{c) $R^2\mathcal{X}_4$ gravity}

In this case, the curvature scale is determined by $1-f_{\infty}-576\lambda_{2,1}f_{\infty}^{4}=0$,
and the functions parameterizing the corrected black hole \reef{bb6} are
\begin{eqnarray}\labell{f1f2}
f_1(z)&=&\frac{432z^3(z^3+z_{\text{H}}^3)}{z_{\text{H}}^6}\, ,\\ \notag
f_2(z)&=&\frac{-144z^3(11z^3-3z_{\text{H}}^3 )}{z_{\text{H}}^6}\,.
\end{eqnarray}
Further, the Hawking temperature becomes
\be \labell{T21}
T=\frac{3}{4\pi z_{\text{H}}}\left(1-144\lambda_{2,1}+\mathcal{O}(\lambda_{2,1}^2)\right)\, ,
\ee
while the entropy density is given by
\be\labell{gramc}
s=\gamma_{c}\,\cse\,T^2 \qquad{\rm with}\ \ \gamma_{c}= 1-576\lambda_{2,1}+\mathcal{O}(\lambda_{2,1}^2)\, .
\ee
Here again, $\gamma_c$ agrees with the analogous factor appearing in the corner coefficient \reef{chargeVW} for $v=2$ and $w=1$. 

\subsubsection*{d) $\mathcal{X}_4^{\,2}$ gravity}

The last nontrivial interaction at fourth order in curvature corresponds to the square of the Gauss-Bonnet density, $\mathcal{X}_4^{\,2}$.  To begin, let us note that interactions of the form $\mathcal{X}_4^{\,w}$ with $w\ge2$ are not topological and do modify the gravitational equations of motion in four dimensions. It is only the linear term, \ie $w=1$ (and $v=0$), which leaves the equations of motion unchanged. 

Now in this case, we have $1-f_{\infty}-96\lambda_{2,1}f_{\infty}^{4}=0$ and 
\begin{eqnarray}\labell{f1f2}
f_1(z)&=&\frac{8z^3(11z^6+z^3z_{\text{H}}^3+z_{\text{H}}^6)}{z_{\text{H}}^9}\, ,\\ \notag
f_2(z)&=&\frac{8z^3(67z^6-83z^3z_{\text{H}}^3+z_{\text{H}}^6)}{z_{\text{H}}^9}\, . \notag
\end{eqnarray}
The Hawking temperature is given by
\be \labell{T21}
T=\frac{3}{4\pi z_{\text{H}}}\left(1-8\lambda_{0,2}+\mathcal{O}(\lambda_{0,2}^2)\right)\,,
\ee
and the thermal entropy density becomes
\be\labell{gramd}
s=\gamma_{d}\,\cse\,T^2\, , \qquad{\rm with}\qquad \gamma_{d}= 1+16\lambda_{0,2}+\mathcal{O}(\lambda_{0,2}^2)\, .
\ee
Here, the factor $\gamma_d$ receives a correction which is first order in $\lambda_{0,2}$ while the corresponding factor in the corner coefficient does not, \eg see eq.~\reef{chacha}. Hence, we have found the first example for which the agreement is broken between the charges defined by the thermal entropy density and by the corner contribution of the entanglement entropy.\\

Gathering together all of  the first order contributions from the new interactions appearing in the fourth-order action (\ref{fo}), we have that the thermal entropy density in the dual boundary theory takes the expected form \reef{thermS} where the corresponding charge takes the form
\be
\cs=\gamma\ \cse
\labell{finalSD}
\ee
where the Einstein result $\cse$ is given in eq.~\reef{chargeSE} and
\be\labell{chachaS}
\gamma=1-24\lambda_1-6\lambda_2+432\lambda_{3,0}+24\lambda_{1,1}-6912\lambda_{4,0}-576\lambda_{2,1}+16\lambda_{0,2}+\mathcal{O}(\lambda^2)\,\, .
\ee
Comparing with eqs.~\reef{finfin} and \reef{chacha} for the corner contribution of the entanglement entropy in the same theories, we see 
\be
\frac{\ka}{\cs}=\frac{9}{8\pi^3}\,\Gamma\!\left({\textstyle \frac{3}{4}}\right)^4\,\left(1-16\lambda_{0,2}+\mathcal{O}(\lambda^2)\right)\,. 
\labell{ratioG}
\ee
That is, the ratio $\ka/\cs$ is independent of most of the additional dimensionless couplings in eq.~\reef{fo} and it would still be given by the same numerical factor found for Einstein gravity in eq.~\reef{ratioE2} for the class of theories with $\lambda_{0,2}=0$.

\subsection{Stress tensor two-point function}\labell{tt}

Let us now turn to the two-point function for the stress tensor, which is particularly interesting since it defines a  central charge for CFT's in any spacetime dimension.  Evaluated in the vacuum, the functional form of this two-point correlator is completely fixed by conformal symmetry and energy conservation, and for a $d$-dimensional CFT, it takes the form \cite{Erdmenger:1996yc,Osborn:1993cr}\footnote{Note that in this section unhatted indices from the beginning of the Latin alphabet run over the $d$-dimensional boundary of AdS$_{d+1}$.}
\be \labell{t2p}
\left\langle\, T_{a b} (x)\, T_{cd}(0)\, \right\rangle=\frac{\ctt}{x^{2d}}\,\mathcal{I}_{ab,cd}(x)\, ,
\ee
where
\be
\mathcal{I}_{ab,cd}(x)\equiv \frac12\left( I_{ac}(x)\,I_{db}(x)+ I_{ad}(x)\,I_{cb}(x)\right)-\frac{1}{d}\,\delta_{ab}\,\delta_{cd}
\labell{beg1}
\ee
and
\be
\labell{beg2}
I_{ab}(x)\equiv \delta_{ab}-2\frac{x_a\,x_b}{x^2}\, .
\ee
Below we will focus on $d=3$ but as remarked above, the above expressions provide a definition of $\ctt$ for CFT's in any spacetime dimension. In particular,  eq.~\reef{t2p} is the standard definition of the central charge $c$ in two-dimensional CFT's, \ie $\ctt=c$, while for four dimensions, $\ctt=40\, c/\pi^4$ where $c$ is the coefficient of the Weyl-squared term in the trace anomaly. 

Of course, in a holographic framework, the stress tensor is dual to the normalizable mode of the metric \cite{Witten:1998qj,Gubser:1998bc} and so evaluating eq.~\reef{t2p} requires determining the two-point boundary correlator of the gravitons in the AdS vacuum. This is a standard calculation in the context of Einstein gravity \cite{hong22,gb2} and one finds for three boundary dimensions
\be
\ctte=\frac{3}{\pi^3}\,\frac{\tL^2}{G}\,.
 \labell{cttE}
\ee 
Once again, we see the ubiquitous factor of $\tL^2/G$ and comparing with the corner coefficient \reef{gammaE}, we have
\be
\labell{ratioE3}
\frac{\kae}{\ctte}=\frac{\pi^2}{6}\,\Gamma\!\left({\textstyle \frac{3}{4}}\right)^4\,.
\ee

In order to investigate how the two-point function (or equivalently the graviton propagator) is modified by the introduction of higher curvature terms in the bulk, let us first recall that generically these new interactions will result in the appearance of higher-order derivatives in the gravitiational equations of motion. Hence the metric will contain additional propagating degrees of freedom beyond the usual massless spin-two graviton. Therefore in a holographic context, the metric will also couple both to the stress tensor and some new tensor operator, which is generically {\it nonunitary}.\footnote{Of course, this is a typical feature of holographic theories with higher curvature interactions in the bulk, but it can be evaded in special cases. For example, $f(R)$ gravity can be re-expressed as Einstein gravity coupled to a scalar field \cite{revfr}. Hence in this case, the additional CFT operator will be a scalar, which can be unitary in the appropriate circumstances. \labell{footing}}  We can understand the latter, \ie that generically the new operator generates negative norm states in the boundary CFT, with the following analogy from \cite{Myers:2010tj}: Consider a massless scalar field in flat space whose equation of motion has been corrected with a fourth-order term, 
\be
\left(\Box+\frac{\lambda}{M^2}\, \Box^2\right)\phi=0\, ,
\ee
where $M^2$ is some high energy scale and $\lambda$, the dimensionless coupling of the higher derivative interaction in the action. Then, the propagator for this field will read
\be
\frac{1}{q^2-\lambda\, q^4/M^2 }=\frac{1}{q^2}-\frac{1}{q^2-M^2/\lambda}\,.
\ee
Here the $q^2=0$ pole will correspond to the usual massless mode, whereas that at $q^2=M^2/\lambda$ is related to a new massive degree of freedom. Regardless of the sign of $\lambda$, the sign of the second term in the propagator above will be negative and so the extra mode is a ghost. Of course, if we are working perturbatively in $\lambda$, these new degrees of freedom appear at very high energy scales. Hence if we should restrict our attention to energies much less than $M/\lambda^{1/2}$, the new scalar ghost will not go on-shell. In the holographic context, the additional ghost modes create negative norm states in the bulk theory and so they must be dual to new nonunitary operators in the boundary theory. Further, let us note that the curvature scale plays the role of the mass above, \ie $L^2\sim 1/M^2$, and so we can expect that the conformal dimension of these operators to be set by the inverse of the gravitational couplings, \ie $\Delta^2\sim1/\lambda$. Hence if we consider the CFT on the background $R\times S^{d-1}$, the new operator would again be associated with high energy states.

The above example also highlights that in a perturbative framework, the extra degrees of freedom are highly suppressed in the vicinity of the physical pole. Hence our strategy in studying the graviton propagator will be to organize the linearized gravitational equations of motion which make this suppression manifest and allow us to easily identify the proper kinetic term of the physical modes. In general, writing out the linearized equations of motion for the graviton would be a very complex task but it can simplified here in two ways, as discussed in \cite{Myers:2010ru}. First, we are interested in the holographic version of eq.~\reef{t2p} which is evaluated in the vacuum and so we need only study the metric fluctuations in the AdS$_4$ background. That is, we consider a perturbed metric: $g_{\mu\nu}= \bar{g}_{\mu\nu}+h_{\mu\nu}$, where $\bar{g}_{\mu\nu}$ is the AdS$_4$ metric (and $h_{\mu\nu}\ll 1$ for all $\mu,\nu=0,1,2,3$). In particular then, the background curvature tensor takes the form $\bar{R}^{\mu\nu}{}_{\sigma\rho}=-1/\tL^2 \left(\delta^\mu{}_\sigma\,\delta^\nu{}_\rho - \delta^\mu{}_\rho\,\delta^\nu{}_\sigma\right)$, which greatly simplifies the form of the linearized equations of motion. That is, they can be expressed entirely in terms of covariant derivatives acting on $h_{\mu\nu}$. In order to further simplify the resulting expressions, which are still rather involved in general, we can use diffeomorphism invariance to choose a convenient gauge. In the following, we restrict ourselves to a transverse traceless gauge,\footnote{Let us comment that in the perturbative framework discussed here, the physical degrees of freedom still correspond to a massless spin-two graviton and so this gauge can still be applied here. Note that the traceless condition eliminates the possibility of identifying new scalar degrees of freedom, \eg as appear in $f(R)$ gravity --- see footnote \ref{footing}. However, these modes are regarded as unphysical with our current perturbative perspective.}  \ie $\bar{\nabla}^{\mu}h_{\mu\nu}=0$ and $\bar{g}^{\mu\nu}h_{\mu\nu}=0$.

With these choices, the linearized Einstein equations become
\be \labell{lineaR}
G^L_{\mu\nu}=-\frac{1}{2}\left[\bar{\Box}+\frac{2}{\tilde{L}^2}\right]h_{\mu\nu}=8\pi G\, {T}_{\mu\nu}\,,
\ee
where $G^L_{\mu\nu}$ denotes the linearized Einstein tensor. We have included the stress tensor ${T}_{\mu\nu}$ for some  additional matter fields to the right-hand side because in the following, it will be important to establish the normalization of Newton's constant, or alternatively of the graviton kinetic term. The linearized equation which results from our complete fourth-order gravity (\ref{fo}) turns out to read\footnote{This result agrees with that found in \cite{Smolic:2013gz} for four-dimensional curvature-squared gravities.}
\be \labell{eomg}
-\frac{\alpha}{2} \left[\bar{\Box}+\frac{2}{\tilde{L}^2}\right]h_{\mu\nu} -
\frac{\lambda_2L^2}{2} \left[\bar{\Box}+\frac{2}{\tilde{L}^2}\right]^2 h_{\mu\nu}=8\pi G \,{T}_{\mu\nu}\, ,
\ee
where $\alpha$ is precisely the constant given by eq.~(\ref{chacha}). Interestingly, none of the higher-order terms considered, except for the $R_{\mu\nu}R^{\mu\nu}$ interaction, produce fourth-order derivatives contributions to the linearized equation for the physical graviton $h_{\mu\nu}$ in the AdS$_4$ background in this gauge, which is a rather striking phenomenom.\footnote{In appendix \ref{appf}, we perform the detailed calculation in a general gauge for $f(R)$ gravity, in which the same structure is found, and show how the fourth-order terms go away in this gauge.} It would be certainly interesting to classify the families of higher-order gravities for which this behaviour is encountered at each order in curvature. We will not pursue such a goal here.

The left-hand side of eq.~(\ref{eomg}) is organized in a way which makes obvious the suppression of the second term in the vicinity of the physical pole, \ie for $(\bar{\Box}+2/{\tilde{L}}^2)h_{\mu\nu}\simeq 0$. However, the higher curvature terms still make their presence felt through the appearance of $\alpha$ which modifies the coefficient of the leading Einstein-like term. As commented above, one can interpret this new coefficient as modifying the normalization of Newton's constant, \ie $G_\mt{eff}=G/\alpha$ or as having modified the normalization of the graviton kinetic term. In any event, the net effect is to modify the previous holographic calculation of the two-point correlator for Einstein gravity by an overall factor of $\alpha$. Hence in the higher curvature theory \reef{fo}, we reproduce the desired expression in eq.~\reef{t2p} where the central charge is now given by
\be\labell{gamend}
\ctt=\alpha\, \ctte=\alpha\, \frac{3}{\pi^3}\,\frac{\tL^2}{G}\, ,
\ee
where again, $\alpha$ is precisely the same constant given by eq.~(\ref{chacha}). Of course, we could also write this expression as $\ctt = 3\tL^2/(\pi^2G_\mt{eff})$, \ie the general result has the same form as that for the Einstein theory except that $G$ is replaced by $G_\mt{eff}$. Therefore, the correction to the central charge appearing in the two-point correlator of the stress tensor \reef{t2p} matches that appearing in the universal corner term. 
Hence all of the higher curvature theories considered here yield the same ratio as in the Einstein theory
\be
\labell{ratioq5}
\frac{\q(\Omega)}{\ctt}=\frac{\qe(\Omega)}{\ctte}
\ee
and in particular, the ratio of charges \reef{ratioE3} is unchanged, \ie
\be
\labell{ratioE4}
\frac{\ka}{\ctt}=\frac{\pi^2}{6}\,\Gamma\!\left({\textstyle \frac{3}{4}}\right)^4\simeq 3.7092\,.
\ee
One might hope that these are universal results extending beyond holography. However, in the discussion section below, we will test this idea by comparing to free field theories. Unfortunately, we find that neither of the above ratios is quite universal but the comparison does show that dividing by $\ctt$ is an interesting way to normalize the corner contribution when comparing different three-dimensional CFT's.

\section{Discussion} \labell{discuss}

In this paper,  we have studied the universal term arising from the presence of corners in the entangling surface for three-dimensional holographic conformal field theories. In general, this coefficient of the logarithmic term in eq.~\reef{one} is a function of the opening angle at the corner $\q(\Omega)$. As we will discuss below, the precise form of this function depends on the details of the underlying CFT, however, as explained in the introduction, this function is constrained to behave as $\q(\Omega)\simeq \ka/\Omega$ in the limit of small opening angles and as $\q(\Omega)\simeq \sigma\,(\Omega-\pi)^2$ in the limit of a nearly smooth entangling surface. 
Hence, eqs.~(\ref{ka}) and \reef{sigmA} define two coefficients, $\ka$ and $\sigma$, which can be used to characterize different CFT's.  Motivated by the idea that the corner contribution provides a useful measure of the number of degrees of freedom in the underlying theory, we referred to these constants as `central charges.' In our holographic calculations, we found that the overall form of $\q(\Omega)$ did not change and so the two charges were simply related in all of holographic models, \ie $\kappa/\sigma=4\,
\Gamma(3/4)^4$. Hence we focus on the small angle charge $\kappa$ in the following discussion. In particular, one goal was to see if this corner charge had a simple relation to any other known `charges,' which provide a similar counting of degrees of freedom and might be accessed with more conventional probes of the theory, or if $\kappa$ is really a distinct quantity. 

Our approach was to study $\kappa$ for an extended holographic model involving higher curvature interactions in the bulk gravity theory, as described in section \ref{seckink}. In particular, we evaluated the corner term for an entangling surface with a sharp corner on the boundary of AdS$_4$, using holographic entanglement entropy \reef{newer}. The final result,
\beq
\ka=\alpha\,\kae\,,  \,\,\, {\rm with}\,\,\,\, \alpha=1-24\lambda_1-6\lambda_2+432\lambda_{3,0}+24\lambda_{1,1}-6912\lambda_{4,0}-576\lambda_{2,1}+\mathcal{O}(\lambda^2)\, ,
\eeq
and
\beq
\kae= \frac{\tilde{L}^2}{2\pi G}\, \Gamma\!\left( {\textstyle \frac{3}{4}}\right)^4\, ,
\eeq
gives $\kappa$ for the broad class of gravitational theories described by the action \reef{fo}. Our general result is proportional to $\tL^2/G$ (\ie the AdS scale squared over Newton's constant) but it is also a function of the eight dimensionless couplings appearing in the action \reef{fo}. Next, in section \ref{witncc}, we evaluated several charges appearing in different physical quantities within the same holographic framework. In particular, we studied the analogous charges appearing in the universal terms in the EE of a strip and of a disk, in the thermal entropy density, and in the two-point function of the holographic stress tensor. All of these measures of degrees of freedom, as well as $\kappa$, are simply proportional to $\tL^2/G$ with Einstein gravity in the bulk and so they can not be distinguished from one another in the corresponding holographic CFT's. However, each of these charges also acquires a distinct dependence on the additional gravitational couplings with higher curvature gravity in the bulk. Our calculations were perturbative in the $\lambda_i$ and hence the results are only linear in these couplings. However, this still allowed us to distinguish the various different charges in the boundary CFT. Hence, this extended holographic model provides an interesting framework to investigate our goal stated above, namely, to determine if the corner charge can be considered distinct or if it has a simple relation to another known central charge.

Of course, we do not have a top-down construction where the action \reef{fo} emerges as the low energy effective action for, \eg some string theory compactification. Rather our perspective is that such extended holographic models provide an interesting framework to test general properties of CFT's, \ie if there are certain properties common to all CFT's then they should be satisfied by the holographic CFT's defined by these models. This approach has found success in a number of interesting contexts, such as the discovery of the F-theorem \cite{Myers:2010xs,Myers:2010tj}. Below, we also look to test a simple conjecture, motivated by our holographic results, with calculations for free massless quantum field theories. 
 
Another caveat in our analysis is that for the generalized Lovelock theories \reef{SWact}, the appropriate gravitational entropy functional to use in evaluating the HEE \reef{newer} is given by eq.~\reef{sw}. Recall that present evidence \cite{trouble} suggests that the general formula for the entropy functional proposed in \cite{Dong} must be further refined for higher curvature theories involving cubic and higher powers of the curvature. However, we argued that the use of eq.~\reef{sw} is well motivated by the somewhat complementary analysis of \cite{Sarkar} examining the second law of black hole thermodynamics in these higher curvature theories --- see also \cite{wall}. However, it would be useful to verify this more directly when a fuller understanding of HEE in higher curvature theories emerges.

A summary of the ratios corresponding to the different charges computed in this paper with respect to $\ka$ can be found in Table \ref{table}.
\begin{center}
\begin{table}[h]
    \begin{tabular}{ | p{3cm} | p{11.5cm} |}
    \hline
    Constant & Ratio  \\ \hline
    Strip HEE  & $\kappa/\ta=1$  \\ \hline
    Disk HEE  & $\ka/c_0=\frac{2}{\pi}\,\Gamma\!\left({\textstyle \frac{3}{4}}\right)^4\,\left(1-2\lgb-24\lambda_{1,1}+288\lambda_{2,1}+96\lambda_{0,2}+\mathcal{O}(\lambda^2) \right)$ \\ \hline
Thermal entropy &$\ka/\cs=\frac{9}{8\pi^3}\,\Gamma\!\left({\textstyle \frac{3}{4}}\right)^4\,\left(1-16\lambda_{0,2}+\mathcal{O}(\lambda^2)\right)$  \\ \hline
$\left< T_{a b}  (x) T_{cd}(0) \right>$ &$\ka/\ctt=\frac{\pi^2}{6}\,\Gamma\!\left({\textstyle \frac{3}{4}}\right)^4$  \\ \hline
\end{tabular}
\caption{Ratios comparing the corner charge $\ka$ with similar physical coefficients.}
\label{table}
\end{table}
\end{center}
\vspace{-0.7cm}

We have seen that our holographic calculations yield $\ka=\ta$, where the latter is the coefficient of the universal term in the EE of a strip, as defined in eq.~(\ref{charstrip}). However, this is a universal result that is expected to hold for any CFT on the basis of a conformal mapping which relates the two entanglement entropy calculations --- see appendix \ref{app4}. Hence this result can be considered a check of our holographic calculations.

On the other hand, the charge $c_0$ corresponding to the universal constant in the EE of a disk is a distinct charge. Of course, the latter is the central charge which decreases monotonically in RG flows, according to the $F$-theorem \cite{Myers:2010xs,Myers:2010tj,Fthem1,Fthem2,proof}. The independence of $\ka$ and $c_0$ is illustrated by eq.~\reef{ratiokc}, which shows that the ratio $\ka/c_0$ depends on $\lgb$, $\lambda_{1,1}$, $\lambda_{2,1}$ and $\lambda_{0,2}$. Hence these two charges depend on the details of the corresponding boundary theories in different ways. Alternatively, the ratio is independent of the remaining four gravitational couplings, $\lambda_{1}$, $\lambda_{2}$, $\lambda_{3,0}$ and $\lambda_{4,0}$. Hence there are also broad classes of theories with the same ratio $\ka/c_0$ but it is not a universal feature common to all CFT's. 

The thermal entropy density for the holographic theories was calculated as the entropy density of the corresponding AdS$_4$ planar black hole. In this case, eq.~\reef{ratioG} shows that $\ka/\cs$ is not universal but only depends on $\lambda_{0,2}$, the coupling for the $\left(\mathcal{X}_4\right)^2$ interaction in eq.~\reef{fo}. However, the fact that this particular example produces a mismatch suggests that this ratio will also depend on other new couplings for more general higher curvature theories. In fact, our findings seem to suggest that the generalized Lovelock theories with $w=0$ or 1 and arbitrary $v$ will respect the agreement between the charges, whereas those with $w\ge 2$ will not. We have explicitly verified that this is the case for $v=1$ and $w=2$.\footnote{That is, $\ka/\cs$ depends on $\lambda_{1,2}$ but $\ka/\ctt$ still takes the standard Einstein value \reef{ratioE3}.}

Eq.~\reef{ratioE4} shows that the ratio $\ka/\ctt$ is the same for all of the holographic theories which we studied, where $\ctt$ is the central charge appearing in the two-point function \reef{t2p} of the stress tensor. Hence eq.~\reef{ratioE4} matches the result \reef{ratioE3} for Einstein gravity with $\ka/\ctt=\pi^2\Gamma\!\left({ \tfrac{3}{4}}\right)^4/6$, at least to first order in the gravitational couplings. It is natural to conjecture that this ratio is a universal quantity for all CFT's, even beyond holography. Some further suggestive results can be found in \cite{Myers:2012vs}, which studied singular entangling surfaces in holographic models in higher dimensions. In particular, the holographic model examined there had Gauss-Bonnet gravity in the bulk and it was found that for an entangling surface with a conical singularity, $\ctt$ controls the coefficient for the universal contribution in the limit of a small opening angle. We will test this simple conjecture below with massless free field theories finding that this ratio is not quite the same in those simple field theories. However, this comparison does support the idea that $\ctt$ provides an interesting normalization of the corner contribution when comparing different CFT's
in three dimensions.

To close let us observe that a consequence of our results is that $\ctt$ and $\cs$ are found to disagree in general for holographic CFT's. Supporting evidence of this disagreement for general holographic theories can be found in \cite{Myers:2010tj}, where it was shown that these two charges are not the same for quasi-topological gravity \cite{Myers:2010ru,Myers:2010jv}.

\subsection{Shape of the extremal surface}
\labell{goliger}

In our holographic investigation of the corner contribution, we found that none of the higher curvature interactions which we studied led to any modification in the functional form of $\q(\Omega)$. Rather it remained exactly the same as in Einstein gravity, \ie  $\q(\Omega)=\alpha\,\qe(\Omega)$ where the constant $\alpha$ is given in eq.~\reef{chacha}. This result is related to the fact that all of the corresponding entropy functionals were extremized by extremal area surfaces in the AdS$_4$ background, just as in Einstein gravity. Further our discussion in section \ref{cs} suggests that this result is {not} simply a consequence of working to first order in a perturbative treatment of the gravitational couplings. Hence one may wonder whether this is a general feature of HEE in the AdS$_4$ vacuum for any higher curvature theory of gravity in the bulk. However, we argue that the latter is, in fact, {\it not} a universal result.

First we observe that the curvature tensor takes the simple form
\be
R_{\mu\nu\rho\sigma}=-\frac1{\tL^2}\left(g_{\mu\rho}\,g_{\nu\sigma}- g_{\mu\sigma}\,g_{\nu\rho}\right)
\labell{curvature}
\ee
in the AdS$_4$ background. Hence as in the examples in section \ref{hcg}, the terms in the entropy functional constructed with background curvatures will reduce to an integral of some constant over the bulk surface $m$, \ie they multiply the Bekenstein-Hawking contribution by some constant factor. Similarly, any terms involving a mixture of background curvatures and extrinsic curvatures will reduce to an integral of some scalar constructed purely from extrinsic curvatures (and possibly derivatives of the extrinsic curvatures). Therefore, we should consider whether in general such extrinsic curvature terms can lead to modifications in the shape of $m$ --- and functional corrections to $\q(\Omega)$, as a consequence. Of course, the intuitive answer, which we confirm below, is that a sufficiently complicated contraction of extrinsic curvatures will have a nontrivial effect on the shape of $m$.  

Following the discussion in section \ref{cs}, we first observe that any term which contains two or more factors of the trace of the extrinsic curvature, \eg $K^{\hat{a}} K^{\hat{a}ij} K^{\hat b}_{ij} K^{\hat b}$,  will always leave the extremal area surface unchanged. The reason is simply that $K^{\hat{a}}=0$ is the equation of motion determining the profile on an extremal area surface. Hence, the variation of a term with two or more factors of $K^{\hat a}$ will produce terms which still contain this factor and so will vanish on any extremal area surface. On the other hand, one might guess that if the term $K^{\hat{a}}_{ij}K^{\hat{a}ij}$ appears in the entropy functional that it will modify the shape of the bulk surface, but we argue that in fact it also leaves the extremal area surface unchanged. This term is actually produced by a curvature squared interaction of the form $R_{\mu\nu\rho\sigma}R^{\mu\nu\rho\sigma}$ \cite{RScalc,squash}. However, this term can be easily rewritten as a linear combination of $R^2$, $R_{\mu\nu}R^{\mu\nu}$ and $\mathcal{X}_4$ interactions, \ie see eq.~\reef{X4}. For a pure AdS$_4$ background, we have argued in section \ref{cs} that extremal area surfaces always extremize the entropy functionals corresponding to each of these three interactions, so the same must be true with $K^{\hat{a}}_{ij}K^{\hat{a}ij}$. However, we find that terms of the form  $\left(K^{\hat{a}}_{ij}K^{\hat{a}ij}\right)^n$ with $ n\geq 2$ are not extremized by the extremal area surface and so we expect contributions of this kind (if they appear in the HEE formula) will modify the functional form of $\qe(\Omega)$. Similarly, cyclic contractions of extrinsic curvatures, \eg $K^{\hat a}_{k_1 k_2}K^{\hat{a} k_2 k_3}\,...\,K^{\hat{d}k_{n-1}k_n}K^{\hat{d}\,k_1}_{k_n}$, would also modify the profile of the bulk surface so they would also change the functional form of the corner function. 

Hence it is relatively simple to find terms which, if they appear in the gravitational entropy functional, would modify the profile of the bulk surface in the calculation of HEE. Hence the general expression for the universal corner term for arbitrary high curvature theories might be expected to take the form
\be \labell{sfinal}
S_{\text{corner}}=-\q(\Omega)\log\left(\frac{H}{\delta} \right)\,,\, \, \text{where}\,\, \, \q(\Omega)=\alpha\,\qe(\Omega)+r(\Omega)\, ,
\ee
where $r(\Omega)$  would be a new function of the opening angle which would depend on some gravitational couplings. If we consider the higher curvature terms as small corrections to Einstein gravity, as for the perturbative calculations in this paper, it should be clear that $r(\Omega)$ would be highly suppressed with respect to the $\qe(\Omega)$ contribution, since it would only start appearing with interactions that are cubic or higher-order in the curvature. On the other hand, as explained in appendix \ref{app4}, even if such functions correct the functional form of $\qe(\Omega)$ for certain higher-order gravities, the small angle behavior of $\q(\Omega)$ is still constrained to take the form
\beq
\lim_{\Omega\to0} \q(\Omega)=\lim_{\Omega\to0} \left(\alpha \,\qe(\Omega)+r(\Omega)\right)= \frac{\kappa}{\Omega}+\cdots\,. \labell{wacwac}
\eeq
Further, as explained in the introduction, we will have
\beq
\lim_{\Omega\to\pi} \q(\Omega)=\lim_{\Omega\to\pi} \left(\alpha \,\qe(\Omega)+r(\Omega)\right)= \sigma\, (\pi-\Omega)^2+\cdots\,, \labell{wacwac2}
\eeq
in the limit of a nearly smooth entangling curve. That is, eqs.~\reef{ka} and \reef{sigmA} will still define the universal corner charges, $\ka$ and $\sigma$, for any holographic theory irrespective of the details of the entropy functional.  However, let us note that for Einstein gravity and for all of the holographic theories studied here, these charges are simply related by $\kae/\sigma_{\ssc E}=4\,\Gamma(3/4)^4$. In general high curvature theories where the corner term is modified as in eq.~\reef{sfinal}, there will be no reason to expect that this simple relation still holds for these two charges.

Of course, we are not at present able to provide an explicit example of a higher curvature interaction which contributes such an `interesting' extrinsic curvature term to the graviational entropy functional. However, in this regard, we are simply restricted by the current limitations in understanding how to construct the entropy functional given a particular interaction in the bulk action \cite{trouble}. Still we do see no reason why these more complicated extrinsic curvature terms can not be produced by sufficiently complicated higher curvature interactions.

\subsection{Comparison with QFT calculations}
\labell{free}

The holographic calculations performed here are expected to produce $\q(\Omega)$ for certain strongly coupled three-dimensional CFT's dual to our bulk gravity theories. On the other hand, similar field theory results are also available for a wide range values of $\Omega$ in the case of a free massless scalar and a free massless fermion \cite{Casini:2006hu,Casini:2009sr,Casini:2008as}.\footnote{Most other  results in the literature, \eg see \cite{2014arXiv1401.3504K,PhysRevB.84.165134,PhysRevLett.110.135702,PhysRevB.86.075106,2013NJPh...15g3048I}, are given only for a particular value of the opening angle, \ie for $\Omega=\pi/2$ which is easily studied on a square lattice.} Further, it was argued \cite{Casini:2006hu,Nishioka:2009un} that the holographic result for the corner contribution $\qe(\Omega)$ with Einstein gravity qualitatively agrees with these free field results. Given how dissimilar the underlying field theories are in this comparison, even a qualitative agreement may seem somewhat surprising. However, recall that the behaviour of $\q(\Omega)$ is fixed on general grounds both for small angles and for $\Omega \simeq \pi$, \ie see eqs.~\reef{ka} and \reef{sigmA}, respectively. Further, given the universal form of $\qe(\Omega)$ at least for the broad range of holographic theories considered in this paper, we find it interesting here to make a quantitative comparison of $\q(\Omega)$ for the holographic and free field theories. In order to make such a comparison, we must start by normalizing $\q(\Omega)$ for the various theories. A convenient choice is to consider $\q(\Omega)/\kappa$ which will then approach  $1/\Omega$ for small angles for any field theory. For all of the holographic theories which we studied, we will have $\qe(\Omega)/\kae$ since the common factor of $\alpha$ in eq.~\reef{finfin} cancels in the ratio. Of course, $\qe(\Omega)$ is determined numerically by evaluating the integrals in eqs.~(\ref{q3}) and (\ref{om}), while $\kae$ is given by eq.~\reef{gammaE}.  The corresponding charges for the free field theories were determined in \cite{Casini:2006hu,Casini:2009sr,Casini:2008as} as
\be
\labell{kasf}
\kappa_{\rm scalar}\simeq 0.0397\qquad{\rm and}\qquad
\kappa_{\rm fermion}\simeq 0.0722\,.
\ee
Now the free field results shown in figures \ref{fixx} and \ref{fiqk} represent Taylor expansions of $\q(\Omega)$ around $\Omega=\pi$ to fourteenth order, which were obtained in \cite{Casini:2009sr,Casini:2008as}. These expansions give a reliable enough approximation for values of the opening angle which are not too small. In particular, the figures also show the lattice results obtained for $\q(\Omega)$ at $\Omega=\pi/4$, $\pi/2$ and $3\pi/4$ in \cite{Casini:2006hu} using the numerical method developed in \cite{Peschel}.
\begin{figure}[h]\centering
   \includegraphics[scale=0.56]{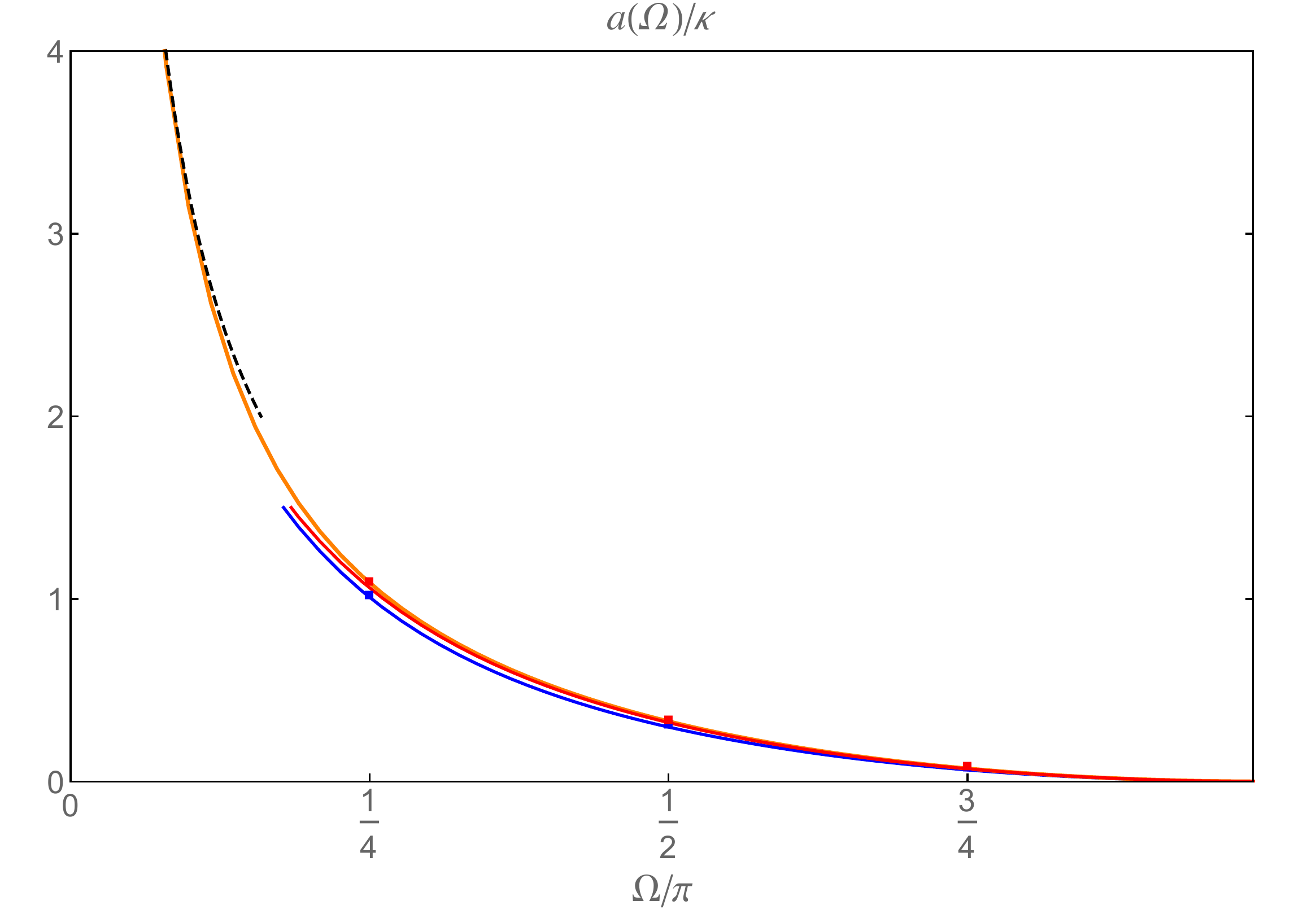}
\caption{(Colour online) We show $\q(\Omega)/\kappa$ for AdS/CFT (orange), a free scalar (blue), a free fermion (red) and the lattice points (squares) obtained numerically for three values of $\Omega$ \cite{Casini:2006hu}. We also include the black dashed curve giving the $1/\Omega$ behavior which all of the functions will approach for small angles.}
 \labell{fixx}
\end{figure}
\begin{figure}[h]\centering
   \includegraphics[scale=0.58]{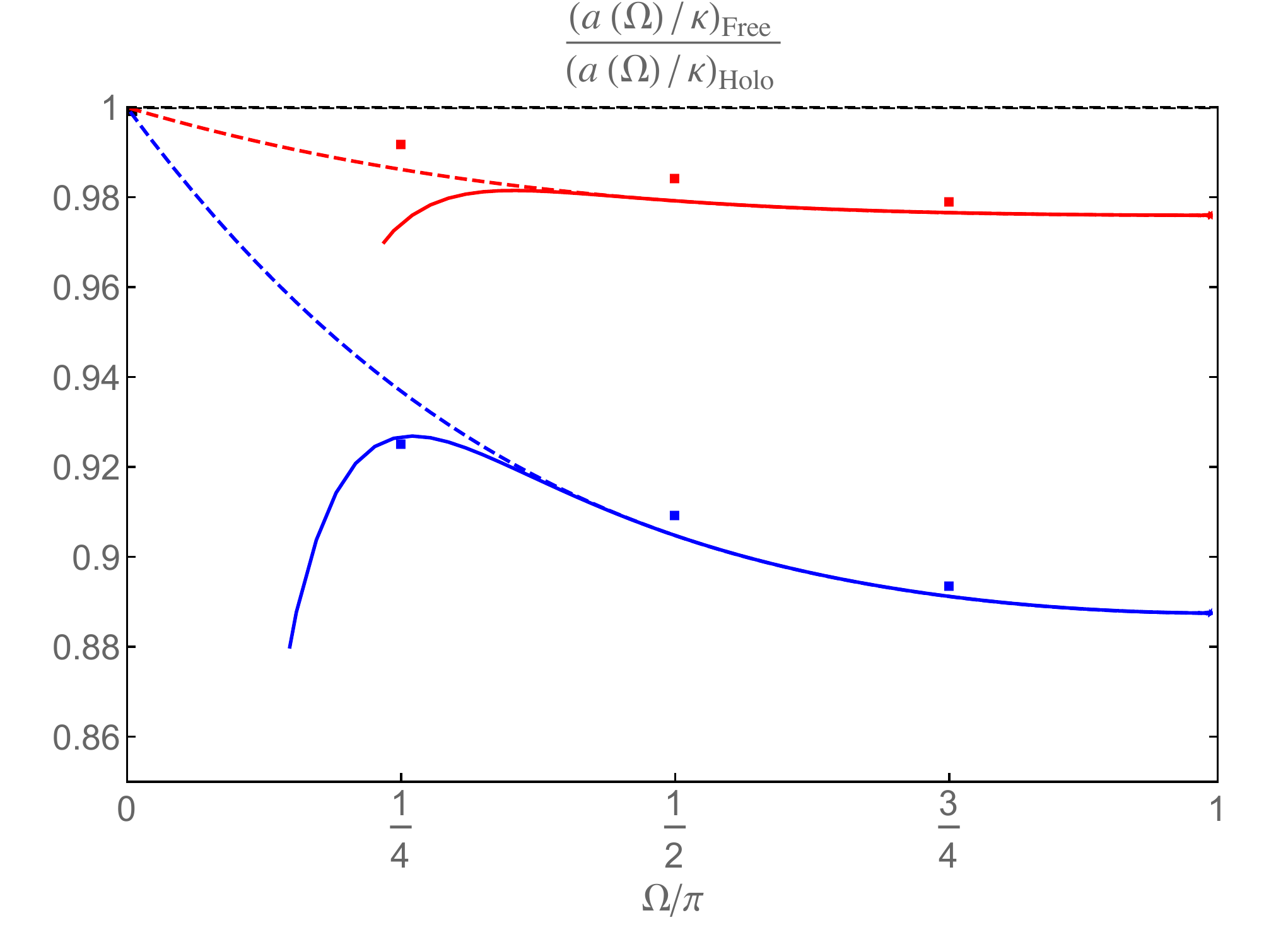}
\caption{(Colour online) We show $(\q(\Omega)/\kappa)_{\rm free}/(\q(\Omega)/\kappa)_{\rm holo}$ both for the free scalar (blue), the free fermion (red) and the corresponding lattice results (squares). We also show the interpolated curves obtained using the 14 coefficients of the Taylor expansions around $\Omega=\pi$ as well as the coefficients $\kappa$ in the small opening angle expansions (dashed blue and red). The black dashed line would correspond to the value for which the ratios are equal. Both theories will in fact approach the black square at the end of this line, \ie at $\Omega=0$, a behaviour that is captured by the interpolated functions.}
 \labell{fiqk}
\end{figure}

In figures \ref{fixx} and \ref{fiqk} we see, first of all, how the Taylor expansions for the free theories are in good agreement with the corresponding lattice results. Hence the red and blue lines in these figures can be reasonably trusted at least for angles larger than $\pi/4$. As we see in figure \ref{fiqk}, the holographic function $\qe(\Omega)/\kae$ turns out to agree with the corresponding free fermion result within a $2\%$ over this whole range where the results are reliable. Similarly, the function for the free scalar deviates from the holographic result by no more than $11\%$ in this range. In the small angle region, the three corner contributions normalized by $\ka$ in figure \ref{fixx} will all approach $1/\Omega$ (shown as the black dashed line). Of course, we only see the latter behaviour is realized for the holographic result, for which we have the exact function over the whole range of $\Omega$. The exact curves for the free scalar (fermion) would lie somewhere in between the black and the blue (red) curves in the intermediate region and so these curves will tend to lie slightly above those obtained with the Taylor series expansion around $\Omega=\pi$. Hence the exact results for the free fields would be in even better agreement with the holographic curve than we have estimated above. Figure \ref{fiqk} is also useful to determine a better estimate of where the Taylor expansions stop being reliable. Focusing on the lattice results in this figure, one might expect that the ratios $(\q/\kappa)_{\rm free}/(\q/\kappa)_{\rm holo}$ for both the scalar and the fermion will decrease monotonically for increasing $\Omega$ over the full range from $\Omega=0$ to $\pi$. This would indicate that the expansions are starting to fail in the vicinity where their slopes become zero, \ie around $\Omega/\pi\sim 0.35$ for the fermion and $\Omega/\pi\sim 0.27$ in the case of the scalar.

As we have seen, the ratio $\ka/\ctt$ equals the Einstein gravity result \reef{ratioE3} for all the higher curvature theories considered here --- at least, for perturbative calculations to linear order in the additional gravitational couplings. However, we might ask if this result applies quite generally for any three-dimensional CFT. Given that for the free field theories, we have at our disposal the values of $\ka$ in eq.~\reef{kasf}, it is interesting to compare these corner charges to the corresponding values of $\ctt$, which can be found in \cite{Osborn:1993cr}: 
\be
\labell{ctsf}
{\ctt}_{\rm \!,\,scalar}= \frac{3}{32\pi^2}\,, \qquad
{\ctt}_{\rm \!,\,fermion}= \frac{3}{16\pi^2}\,.
\ee
Hence the ratios become:
\be
\labell{ratioEsD}
\frac{\kappa}{\ctt}\Big|_{\rm holo}\simeq 3.7092\,,\qquad
\frac{\kappa}{\ctt}\Big|_{\rm scalar}\simeq 4.17945\,,\qquad
\frac{\kappa}{\ctt}\Big|_{\rm fermion}\simeq 3.8005\,.
\ee
All of these ratios are rather close to each other but we do not have precise agreement. In particular, the fermion result differs from the holographic one by approximately $2.5\%$ whereas the scalar ratio is off by approximately $13\%$. Of course, an open question which remains is whether this ratio is a universal quantity for all holographic theories, however, we can only begin to address this question when a better understanding is established for holographic entanglement entropy in general higher curvature theories.

In fact, it was not only the ratio $\ka/\ctt$ but rather the entire function $\q(\Omega)/\ctt$ which was universal for all our higher curvature theories. Hence, even though we found that the universality of $\ka/\ctt$ did not extend beyond holographic CFT's, we may ask more broadly if there are any features of the corner contribution which are universal for general three-dimensional CFT's. Hence in figure \ref{fiqct}, we plot $(\q(\Omega)/\ctt)_{\rm free}/(\q(\Omega)/\ctt)_{\rm holo}$ for the free scalar (blue) and the free fermion (red). The figure also includes the corresponding lattice points\footnote{Although, the Taylor expansions and the lattice points seem to differ here, they are actually in good agreement and it is just that the vertical scale has been expanded here. In particular, the disagreement is less than approximately $2.5\%$ in all cases.} as well as the points at  $\Omega/\pi=0$, which correspond to $(\kappa/\ctt)_{\rm free}/(\kappa/\ctt)_{\rm holo}$. As can be expected from figures \ref{fixx} and \ref{fiqk}, we see that in general the corner contribution evolves slightly differently for the three cases as  $\Omega$ runs from 0 to $\pi$. The ratios plotted in figure \ref{fiqct} are essentially the same in figure \ref{fiqk} except that we have changed the normalization by considering $\q(\Omega)/\ctt$ rather than $\q(\Omega)/\ka$. Hence again, the both ratios in the new figure seem to be monotionically decreasing starting from $(\kappa/\ctt)_{\rm free}/(\kappa/\ctt)_{\rm holo}$ at $\Omega=0$ --- see eq.~(\ref{ratioEsD}). The remarkable feature in figure \ref{fiqct} is that both curves seem to reach precisely 1 at $\Omega=\pi$. That is, it appears that the ratio $\sigma/\ctt$ is equal for the two field theories and for our holographic theories!
\begin{figure}[h]\centering 
   \includegraphics[scale=0.58]{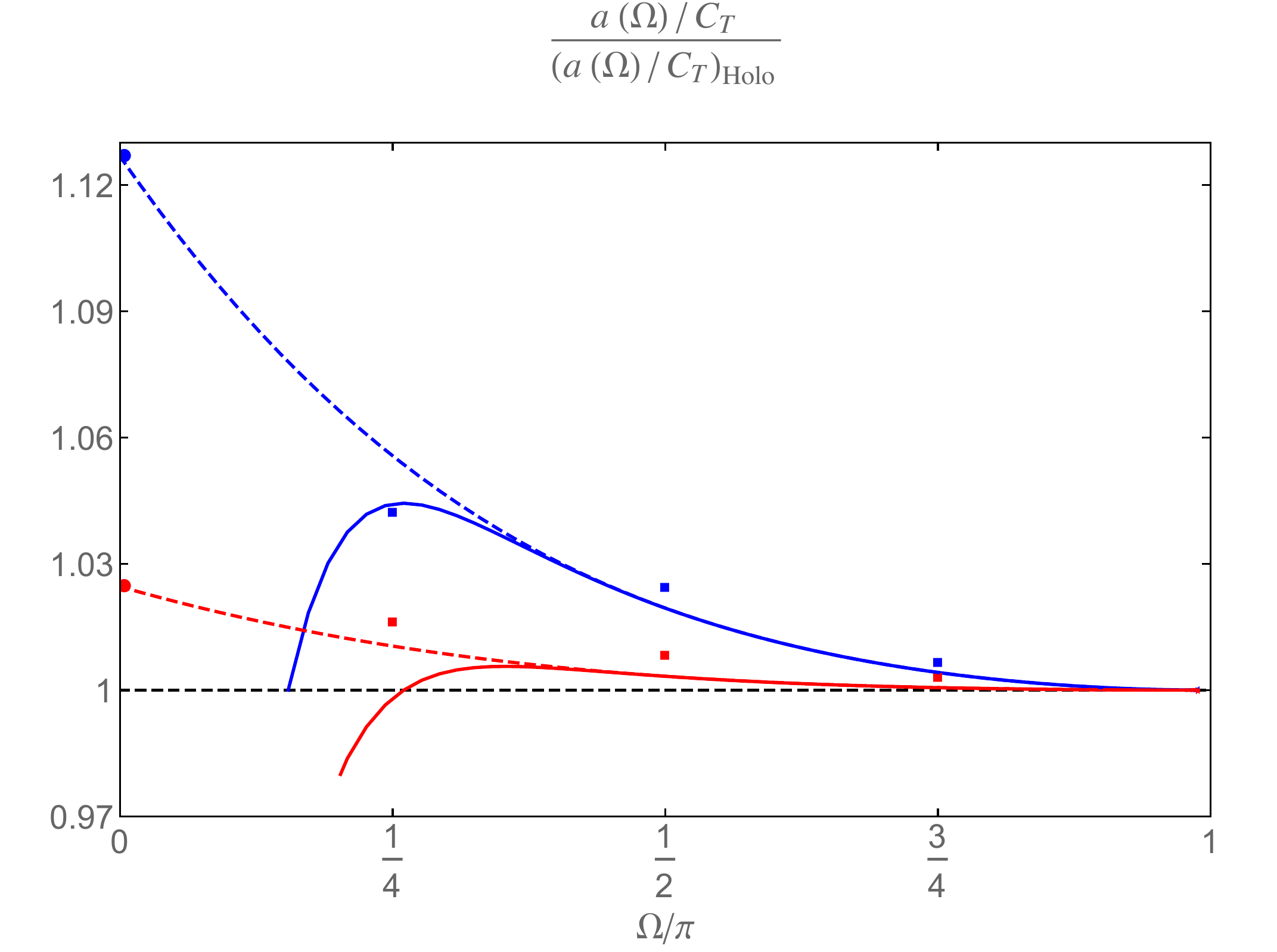}
\caption{(Colour online) We show $(\q(\Omega)/\ctt)_{\rm free}/(\q(\Omega)/\ctt)_{\rm holo}$ both for the free scalar (blue), the free fermion (red) and the lattice points (squares). We also include the interpolated curves obtained using the 14 coefficients of the Taylor expansions around $\Omega=\pi$ as well as the coefficients $\kappa$ in the small opening angle expansions (dashed blue and red). The black dashed line would correspond to the value for which the ratios equal $1$. The dots in blue and red at $\Omega=0$ correspond to the small angle values of the ratios, namely $(\kappa/\ctt)_{\rm free}/(\kappa/\ctt)_{\rm holo}$.}
\labell{fiqct}
\end{figure}

Recall that we argued the behavior of $\q(\Omega)$ was constrained for general CFT's near $\Omega=\pi$ and eq.~\reef{sigmA} defined the charge $\sigma$ with $\q(\Omega)\simeq  \sigma\, (\pi-\Omega)^2 +\cdots$.
In particular, we found in eq.~(\ref{lim2}) that for Einstein gravity
\be
\sigma_{\mt E}=\frac{\tilde{L}^2}{8\pi G}\, ,\labell{gorgon}
\ee
and so given the universal form of $\q(\Omega)$ for all our holographic theories in eq.~\reef{finfin}, we have 
\be
\sigma=\alpha\, \sigma_{\ssc E}\, ,\labell{gorgon2}
\ee
with $\alpha$ given again by eq.~(\ref{chacha}). Further in all of our holographic theories, we also have a fixed ratio:
\be
\frac{\sigma}{\ctt}=\frac{\sigma_{\ssc E}}{\ctte}=\frac{\pi^2}{24}\simeq 0.411234\ .
\labell{ratiosig}
\ee
We can easily compare this result with the ratio $\sigma/\ctt$ for the free conformal scalar and the massless fermion, since $\sigma$ is simply the first nonvanishing coefficient in the Taylor expansions presented in \cite{Casini:2006hu,Casini:2009sr,Casini:2008as}, and the corresponding values are
\be
\labell{kasfx}
\sigma_{\rm scalar}\simeq 0.0039063\ ,\qquad{\rm and}\qquad
\sigma_{\rm fermion}\simeq 0.0078125\ .
\ee
Hence using the values of $\ctt$ given in eq.~\reef{ctsf}, the desired ratios become
\be
\frac{\sigma}{\ctt}\Big|_{\rm scalar}\simeq 0.411235\ , \qquad{\rm and}\qquad
\frac{\sigma}{\ctt}\Big|_{\rm fermion}\simeq 0.411234\ .
\labell{ratiosig2}
\ee
Hence as expected from figure \ref{fiqct}, the free field ratios show a striking agreement with the holographic result, \ie they agree with a precision of better than $0.0003 \%$! We might keep in mind that while the free field values for $\ctt$ in eq.~\reef{ctsf} are exact, the corresponding values of $\sigma$ in eq.~\reef{kasfx} are only the approximate results of a numerical computation \cite{Casini:2006hu,Casini:2009sr,Casini:2008as}. Hence the precision of the agreement between eqs.~\reef{ratiosig} and \reef{ratiosig2} is as good as could be expected. 

These observations, originally made in \cite{BuenoMyersWill}, led us to conjecture there that the ratio $\sigma/\ctt$ is in fact a universal constant for all three-dimensional CFT's, \ie
\be \label{unisc}
\frac{\sigma}{\ctt}=\frac{\pi^2}{24}
\ee
for general conformal field theories in three dimensions. This conjecture can be used to predict the exact values of 
$\sigma_{\rm scalar}$ and $\sigma_{\rm fermion}$,
\be
\labell{kasfpredi}
\sigma_{\rm scalar}=\frac{1}{256}\,,\qquad{\rm and}\qquad
\sigma_{\rm fermion}=\frac{1}{128}\,.
\ee
Of course, these values match the results shown in eq.~(\ref{kasf}) within the accuracy limits set by the calculations in \cite{Casini:2006hu,Casini:2009sr,Casini:2008as}. However, we can improve these results by going back to the original free field computations and evaluating the required integrals with an improved accuracy. The required calculations are described in appendix \ref{integrals} and we find that the agreement between our prediction for $\sigma_{\rm scalar}$ and $\sigma_{\rm fermion}$, given by eq.~(\ref{kasfpredi}), and the previous calculations for the free field results can be extended to an accuracy of one part in $10^{12}$.\footnote{Remarkably, after the first version of this paper appeared in the arXiv, these integrals were evaluated analytically \cite{Henriette}, confirming the results predicted by the conjecture in eq.~\reef{kasfpredi}.} We emphasize the required integrals (\ref{fun1}) and (\ref{fun2}) are extremely complicated and they are not even similar. Yet they seem to conspire to produce the simple rational numbers \reef{kasfpredi} predicted by holography. As discussed in \cite{BuenoMyersWill}, we feel this is striking evidence in favour of our new conjecture above!

While we have found that eq.~\reef{unisc} applies both for the holographic theories and for free CFT's, it would of course be interesting to extend these calculations to other three-dimensional CFT's. For example, one might consider the $\mathcal{N}=2$ critical Wess-Zumino model\footnote{We thank Nikolay Bobev for useful comments on this theory.}. For this theory, $\ctt$ has been computed exactly using localization in \cite{Nishioka:2013gza}, finding
\begin{eqnarray}\hspace{-0.5cm}
{\ctt}_{\rm \!,\,\mathcal{N}=2\, WZ}&=&\int_0^{\infty}\frac{dx}{\pi^4}\,\left[\left(\frac{1}{x^2}-\frac{\cosh(2x/3)}{\sinh^2(x)}\right)+\frac{3(\sinh(2x)-2x)\sinh(2x/3)}{2\sinh^4(x)} \right]\, ,
\end{eqnarray}
which yields
\beq
{\ctt}_{\rm \!,\,\mathcal{N}=2\, WZ} \simeq 0.02761450054158\,.
\eeq
This same numerical value was also recently reproduced using the conformal bootstrap \cite{Bobev:2015vsa,Bobev:2015jxa}.\footnote{Note that in the conventions of \cite{Bobev:2015jxa}, $\ctt$ differs from ours by a factor $16\pi^2$ in $d=3$, \eg ${\ctt}_{\rm \!,\,scalar}=3/2$ in the conventions of \cite{Bobev:2015jxa}. Our conventions are, however, the same as in \cite{Nishioka:2013gza}.} Our conjecture \reef{unisc} would predict the corresponding value of ${\sigma}$ as
\beq
{\sigma}_{\rm \!\,\mathcal{N}=2\, WZ}\simeq 0.011356008\, .
\eeq
Computing this quantity (or the analogous ones in other CFT's for which $\ctt$ is known, such as the $O(N)$ models) would provide a strong test for this conjecture \cite{BuenoMyersWill}.

Let us also observe that the higher order coefficients in the Taylor expansions around $\theta=\pi$ \cite{Casini:2008as,Casini:2006hu,Casini:2009sr} for the free scalar and the free fermion do not seem to exhibit any similar universal behaviour. In particular, these coefficients do not seem to be related in any simple way between the two theories, in contrast to the simple relation $\sigma_{\rm fermion}=2\,\sigma_{\rm scalar}$ above. We might mention that it was already observed in \cite{Casini:2008as,Casini:2006hu,Casini:2009sr} that their numerical results for these coefficients only seemed to differ by a factor of two, but no explanation was given. According to the conjecture \reef{kasfpredi}, the reason comes simply from the well-known result that ${\ctt}_{\rm \!,\,fermion}=2\,{\ctt}_{\rm \!,\,scalar}$ 
 \cite{Osborn:1993cr}.

More generally, the comparison shown in figure \ref{fiqct} illustrates that $\ctt$ provides a useful normalization of the corner contribution when comparing results for different CFT's in three dimensions. In \cite{BuenoMyersWill}, we also considered the Wilson-Fisher fixed points of the $O(N)$ models with $N=1,2,3$. In this case, numerical results $\q(\Omega=\pi/2)$ were available from state of the art numerical simulations of lattice Hamiltonians with the corresponding 
quantum critical points \cite{PhysRevLett.110.135702,pitch,2014arXiv1401.3504K}, while conformal bootstrap methods were recently used to determine $\ctt$ with great accuracy for these theories \cite{Kos:2013tga}.  The agreement with the holographic value for $\q(\Omega=\pi/2)/\ctt$ was better than $12\%$ in all three cases \cite{BuenoMyersWill}.

Beyond pointing out a useful normalization by which the corner term for different CFT's can be compared, the holographic expression for $\q(\Omega)/\ctt$ seems to provide a good benchmark with which to compare the analogous results for general three-dimensional CFT's. As discussed above, we found a surprising agreement between the holographic results and those for both free and strongly interacting CFT's. Of course, it would be interesting to extend these comparisons to other three-dimensional CFT's. For example, one might consider the $\mathcal{N}=2$ critical Wess-Zumino model discussed above.

A further suggestive observation is the holographic result provides the smallest values of $\q(\Omega)/\ctt$, \eg in figure \ref{fiqct}. Hence it would be natural to investigate if  the  holographic result is a universal lower bound for any three-dimensional CFT. This conjecture would be similar to the celebrated KSS conjecture that $\eta/s=1/(4\pi)$ represents an absolute lower bound for any relativistic quantum field theory \cite{Kovtun:2004de}. Of course, this bound was found by investigating holographic CFT's dual to Einstein gravity, but the appearance of higher curvature interactions in the bulk could produce violations of the conjectured bound \cite{steve2}. In contrast, our holographic analysis here shows that  $\q(\Omega)/\ctt$ remains unaffected by a broad class of higher curvature terms. This provides a good motivation for further study of the issues surrounding the shape of the extremal surface appearing in the holographic calculation of $\q(\Omega)$. Certainly, if the corresponding bulk surface is no longer the same as in Einstein gravity, this would modify the 
functional form of $\q(\Omega)$ and hence the lower bound might be violated for some values of the bulk couplings. 

We note that the conjectured universality of $\sigma/\ctt$ is a rather striking result since $\sigma$ characterizes the EE, which can generally be regarded as a nonlocal quantity, while $\ctt$ is defined by a local correlation function \reef{t2p}. However, we expect \cite{prep1} that the universal ratio in eq.~\reef{unisc} can be derived using the techniques developed in \cite{vlad4}, which examine changes in the EE induced by small perturbations of the geometry and couplings. In this situation,  it is clear that these small variations of the EE are indeed controlled by local correlators. To conclude, let us add that holographic calculations suggest that similar universal behaviour also arises in higher dimensions \cite{prep2}. In particular, we can use the holographic results for the quadratic correction to the universal term arising from deformations of spherical entangling surfaces obtained in \cite{mark}.

\acknowledgments
We wish to thank Horacio Casini for providing us with the free field results and lattice data points for $\q(\Omega)$, which were used in figures \ref{fixx}, \ref{fiqk} and \ref{fiqct} as well as the explicit integrals for $\sigma_{\rm scalar}$ and $\sigma_{\rm fermion}$  in appendix \ref{integrals}. We are also thankful to Nikolay Bobev, Horacio Casini, Ana Cueto, Dami\'an Galante, Tarun Grover, Kris Jensen, Ann B. Kallin, \'Oscar Lasso, Patrick Meessen, Roger Melko, M\'ark Mezei, Tom\'as Ort\'in, Pedro F. Ram\'irez, Misha Smolkin, Antony Speranza, Marika Taylor and especially William Witczak-Krempa for useful discussions and comments. PB thanks Perimeter's Visiting Graduate Fellows Program and Perimeter Institute, where most of the project was carried out, for its kind hospitality. Research at Perimeter Institute is supported by the Government of Canada through Industry Canada and by the Province of Ontario through the Ministry of Research \& Innovation. Research at IFT is supported by the Spanish MINECO's \emph{Centro de Excelencia Severo Ochoa} Programme under grant SEV-2012-0249. The work of PB has also been supported by the JAE-predoc grant JAEPre 2011 00452 and partially by the Spanish Ministry of Science and Education grant FPA2012-35043-C02-01, the \emph{Comunidad de Madrid} grant HEPHACOS S2009ESP-1473, and the Spanish Consolider-Ingenio 2010 program CPAN CSD2007-00042.  RCM acknowledges support from an NSERC Discovery grant and funding from the Canadian Institute for Advanced Research.

\appendix

\section{Conventions and notation}\labell{app1}

Here we outline our conventions and notation for the calculations in sections \ref{seckink} and \ref{witncc}. Greek indices run over the entire AdS$_4$ background, whereas Latin letters from the \emph{second} half of the alphabet $i,j,...$ represent directions along the extremal surface $m$. Here $m$ is a (co)dimension-two bulk surface with a pair of independent orthonormal vectors orthogonal to it $n^{\mu}_{\hat{a}}$ ($\hat{a}=\hat{1},\hat{2}$), where the hatted indices from the beginning of the Latin alphabet denote tangent indices in the transverse space, so that $\delta_{\hat{a}\hat{b}}=n^{\mu}_{\hat{a}}n^{\nu}_{\hat{b}}g_{\mu\nu}$. Tangent vectors to $m$ are defined in the usual way as $t_{i}^{\mu}\equiv\partial x^{\mu}/\partial y^i$, being $x^{\mu}$ and $y^i$ coordinates in the full AdS$_4$ background and along the surface, respectively. The corresponding induced metric on the surface is thus given by $\gamma_{ij}\equiv t^{\mu}_it^{\nu}_jg_{\mu\nu}$ (and its determinant $\det \gamma_{ij}\equiv \gamma$). The extrinsic curvatures associated to the two normal vectors  $n^{\mu}_{\hat{a}}$ are given by $K^{\hat{a}}_{ij}\equiv t^{\mu}_it^{\nu}_j\nabla_{\mu}n_{\nu}^{\hat{a}}$, where $\nabla_{\mu}$ is the covariant derivative compatible with $g_{\mu\nu}.$ Also, we use $K^{\hat{a}}$ to denote the trace of each extrinsic curvature defined through $K^{\hat{a}}\equiv \gamma^{ij}K^{\hat{a}}_{ij}$. Finally, with $({K^{\hat{a}}})^2$, we mean the sum of the squares of the two extrinsic curvatures: $({K^{\hat{a}}})^2\equiv K^{\hat{a}}K^{\hat{b}}\, \delta_{\hat{a}\hat{b}}$. The transverse metric can be defined as $g^\perp_{\mu\nu}\equiv n^{\mu}_{\hat{a}}n^{\nu}_{\hat{b}} \delta^{\hat{a}\hat{b}}$, and allows us to project bulk tensors in the transverse directions, \eg $R^{\hat{a}}{}_{\hat{a}}\equiv g^\perp{}^{\mu\nu}R_{\mu\nu}$.

In the calculations of the corner contribution in section \ref{seckink}, we write Euclidean AdS$_4$ in Poincar\'e coordinates as
\begin{equation}\labell{ads4x}
ds^2=\frac{\tilde{L}^2}{z^2}\left(dz^2+d\te^2 +d\rho^2+\rho^2d\theta^2  \right)\, .
\end{equation}
The induced metric on surfaces $m$ parametrized as $t_E=0$, $z=\rho\, h(\theta)$, such as those suitable for the entangling surface with a corner, reads
\begin{equation}
ds^2_{m}=\frac{\tilde{L}^2}{\rho^2}\left(1+\frac{1}{h^2} \right)d\rho^2+\frac{\tilde{L}^2}{h^2}\left(1+\dot{h}^2 \right)d\theta^2+\frac{2\tilde{L}^2\dot{h}}{\rho\, h}d\rho \,d\theta \, ,\labell{indmee2}
\end{equation}
where $\dot{h}(\theta)\equiv \partial_{\theta} h$. From the above, one finds
\begin{equation}\labell{gammai}
\sqrt{\gamma}=\frac{\tilde{L}^2}{\rho h^2}\sqrt{1+h^2+\dot{h}^2}\, .
\end{equation}
The resulting orthonormal vectors orthogonal to the surface read 
\begin{eqnarray}
 n_{\hat{1}}&=&\frac{z}{\tilde{L}}\,\partial_t\, , \\ 
n_{\hat{2}}&=&\frac{z}{\tilde{L}\sqrt{1+h^2+\dot{h}^{2}}}\left(\partial_z-h\,\partial_{\rho}-\frac{\dot{h}}{\rho}\, \partial_{\theta} \right)\, .
\end{eqnarray}
 For our pure AdS$_4$ background, we find the following expression for the projection of the Ricci tensor appearing in eq. (\ref{see3})
\begin{equation}
R^{\hat{a}}{}_{\hat{a}}=g^\perp{}^{\mu\nu}R_{\mu\nu}=-6/{\tilde{L}}^2\, .
\end{equation}
The extrinsic curvature associated to $n_{\hat{1}}$ vanishes, whereas that corresponding to $n_{\hat{2}}$ turns out to be
\begin{equation}
K^{\hat{2}}_{ij}=\left(
\begin{array}{cc}
 -\frac{\tilde{L} \left(h^2+1\right)}{\rho ^2 h^2 \sqrt{h^2+\dot{h}^{ 2}+1}} & -\frac{\tilde{L}\dot{h}}{\rho  h \sqrt{h^2+\dot{h}^{ 2}+1}} \\
 -\frac{\tilde{L}\dot{h}}{\rho  h \sqrt{h^2+\dot{h}^{ 2}+1}}  & -\frac{\tilde{L} \left(h^2+\ddot{h} h+\dot{h}^{ 2}+1\right)}{h^2 \sqrt{h(\theta )^2+\dot{h}^{ 2}+1}} \\
\end{array}
\right)\, .
\end{equation}
From this we can easily obtain the contraction appearing in eq. (\ref{see3})
\begin{equation}
\left({K^{\hat{a}}}\right)^2= \frac{\left[2+3 h^2+h^4+2\dot{h}^2+h(1+h^2)\ddot{h} \right]^2}{\tilde{L}^2\left(1+h^2+\dot{h}^2\right)^3}\, .
\end{equation}
Finally, the intrinsic Ricci scalar evaluated with the metric $\gamma_{ij}$ on the bulk surface reads
\begin{equation}
\labell{RSC}
 \mathcal{R}=\frac{2(-(1+2h^2)\dot{h}^2-\dot{h}^4+(h+h^3)\ddot{h})}{\tilde{L}^2(1+h^2+\dot{h}^2)^2}\, .
\end{equation}
From the above expression and eq.~(\ref{gammai}), it is straightforward to verify that the product $\sqrt{\gamma}\,\mathcal R$ is a total derivative. Indeed, we find
\begin{equation}
\labell{gR}
\sqrt{\gamma}\, \mathcal{R}=\frac{2(-(1+2h^2)\dot{h}^2-\dot{h}^4+(h+h^3)\ddot{h})}{\rho h^2(1+h^2+\dot{h}^2)^{3/2}}=\frac{d}{d\theta}\left[\frac{2}{\rho}\,\frac{\dot{h}}{ h \sqrt{1+h^2+\dot{h}^2}} \right]\, .
\end{equation}

\section{From the corner to the strip}\labell{app4}

As we used in the main text, the small angle limit of $\q(\Omega)$ defines a universal charge $\ka$, which can be used to distinguish different CFT's. The form of eq.~(\ref{ka}) is fixed for general theories due to the existence of a conformal map relating the corner geometry to a strip. This mapping is discussed in detail in Appendix A of \cite{Myers:2012vs} and we only review the salient points here. As a consequence of this mapping,  the expressions for the universal terms in the entanglement entropy match for both geometries, at least in the limit of small $\Omega$ or a narrow strip width. However, as we will see below, this mapping does not fix the form of $\q(\Omega)$ over the entire range of the opening angle.

Let us now describe the conformal mapping: Let the CFT be defined in the background geometry which is simply $\mathbb{R}^3$, with the coordinates used in section \ref{seckink},
\be\labell{r3}
ds^2=dt^2_E+d\rho^2+\rho^2d\theta^2\, .
\ee
If we make the coordinate transformation, $t_E=r \cos\xi$ and $\rho=r \sin \xi$, the line element above becomes
\be\labell{round}
ds^2=dr^2+r^2\left(d\xi^2+\sin^2\!\xi\, d\theta^2\right)\, .
\ee
Next we make the coordinate change $r=L\, e^{Y/L}$ and remove the overall factor $e^{2Y/L}$ with a Weyl transformation, to find the geometry
\be\labell{kinkstrip}
ds^2=dY^2+L^2\left(d\xi^2+\sin^2\!\xi\, d\theta^2\right)\, ,
\ee
with $Y\in \left(-\infty,+\infty \right)$. Of course, this conformal transformation is the usual exponential map which takes $\mathbb{R}^3$ to $\mathbb{R}\times S^2$. 

The corner region for which we calculated the entanglement entropy in section \ref{seckink} was defined in the original coordinates \reef{r3} as $V= \left\{\te=0, \rho>0, |\theta|\le  \Omega/2  \right\}$ and so in terms of the polar coordinates \reef{round}, this region becomes $V= \left\{r >0, \xi=\pi/2, |\theta|\le  \Omega/2  \right\}$. 
Finally in the cylindrical background \reef{kinkstrip}, the corner region is mapped to an infinite strip: $V= \left\{Y\in\left(-\infty,+\infty \right), \xi=\pi/2, |\theta|\le  \Omega/2  \right\}$. In this geometry, the density matrix would be represented by a path integral of the CFT over the cylinder with open boundary conditions imposed along the strip, \ie on surfaces just above and below $\xi=\pi/2$, along the entire length of $Y$ and in the range $|\theta|\le  \Omega/2$. Hence the entire entanglement entropy \reef{one}, including both the universal and nonuniversal contributions, for the corner geometry in $\mathbb{R}^3$ is readily related to that for the strip in the cylinder geometry $\mathbb{R}\times S^2$, as discussed in \cite{Myers:2012vs}. However, we would like instead to relate the entanglement entropy of the corner region to that of a strip in flat space $\mathbb{R}^3$, as was discussed in section \ref{belt}. This is where the  limit of small opening angle becomes important.  When $\Omega\ll 1$, the separation between both sides of the strip is much smaller than the size of the sphere and the local radius of curvature, \ie $\ell\equiv L\,\Omega\ll L$. Hence the latter scale is negligible and to leading order the entanglement entropy resembles that for a strip in flat space, \ie
\be\labell{strip2}
\see = c_1 \, \frac{2(Y_+-Y_-)}{\delta} - \ta\, \frac{Y_+-Y_-}{\ell}+\cO(\delta/L,\ell/L)
\ee
where $Y_+$ and $Y_-$ are regulator scales introduced to cut-off the length of the strip in the positive and negative $Y$ directions \cite{Myers:2012vs} --- compare to eq.~\reef{strip}. Given the preceding transformations, we see that the universal contribution (proportional to $\ta$) is mapped to
\be\labell{strop3}
S_\mt{univ}= -\frac{\ta}{\Omega}\,\log\left(\frac{\rho_{\ssc max}}{ \rho_{\ssc min}}\right)=-\frac{\ta}{\Omega}\,\log\left(\frac{H}{ \delta}\right)\,,
\ee
where we have made the natural substitutions: $\rho_{\ssc max}=H$ and $\rho_{\ssc min}=\delta$. We emphasize that this expression only applies for $\Omega\ll 1$ and hence we have recovered eq.~\reef{ka} for the corner contribution with $\ka=\ta$. 

Let us add that the coordinate transformation in the bulk geometry implementing the conformal mapping between the two boundary metrics (\ref{r3}) and (\ref{kinkstrip}) can be found as follows: The AdS$_4$ geometry can be described as a hyperbola embedded in the five-dimensional Minkowski space
\be\labell{mink5}
ds^2=-dU^2+dV^2+dR^2+R^2\, d\Omega^2_2\, .
\ee
AdS$_4$ is defined now as the subspace
\be \labell{constrat}
-U^2+V^2+R^2=-L^2\, .
\ee
This constraint can be solved writing $R=rL/z$, $U+V=L^2/z$, $U-V=z+r^2/z$, and the induced metric on the hyperbola reduces to the Poincar\'e coordinates on AdS$_4$, given in eq.~(\ref{ads4}). On the other hand, the constraint (\ref{constrat}) is also satisfied by $U=\sqrt{R^2+L^2}\,\cosh(Y/L)$, $V=\sqrt{R^2+L^2}\,\sinh(Y/L)$, in which case the induced metric becomes
\be\labell{ind}
ds^2=\frac{dR^2}{1+\frac{R^2}{L^2} }+\left(1+\frac{R^2}{L^2} \right) dY^2+R^2 \left(d\xi^2+\sin^2\!\xi\, d\theta^2\right)\, ,
\ee
which is the AdS$_4$ geometry in global coordinates. Stripping off a scale factor of $R^2/L^2$ at large radius, the resulting boundary metric matches that in eq.~\reef{kinkstrip}. These bulk coordinates can be used to compute the HEE for the kink in essentially the same way as the calculation of section \ref{seckink}.

\section{$f(R)$ gravity}\labell{appf}

We parmeterize our general $f(R)$ gravity action \cite{revfr} as
\begin{equation}\label{fr}
I_{f(R)}=\frac{1}{16\pi G}\int d^4x\sqrt{g}\left[\frac{6}{L^2} +R+\lhat \,f(R)\right] \, ,
\end{equation}
where we have made the cosmological constant and the Einstein term explict. We have also introduced a dimensionless coupling $\lhat$ as a useful device to indicate the combined strength of the higher curvature contributions in the following. The function $f(R)$ can be a general function of the Ricci scalar, which has a Taylor series expansion beginning at order $R^2$ or higher. Our perspective is that $f(R)$ is parameterized by various dimensionless couplings and the necessary dimensions are provided by the cosmological constant scale $L$. For example, we would incorporate the three Ricci scalar terms in the action \reef{fo} as
\be\labell{examex}
\lhat\,f(R)=L^2\,\lambda_1\,R^2 +L^4\,\lambda_{3,0}\,R^3+L^6\,\lambda_{4,0}\,R^4 \,.
\ee
In this simple class of theories, the gravitational entropy functional is simply given by the Wald entropy \cite{Sarkar,razor}, \ie
\begin{eqnarray}
\labell{seefr}
 S_{f(R)}&=&\frac{1}{4G} \int_{m} \!\!d^2y\, \sqrt{\gamma}\, \left[1+\lhat f^{\prime}(R)\right]  \, ,
\end{eqnarray}
where $f'(R)=\partial f(R)/\partial R$. For our pure AdS$_4$ background, $f^{\prime}(R)$ will be just a constant, with $R=-12/\tL^2\equiv\bar R$ where 
\be\labell{racket}
\frac{1}{L^2}=\frac1{\tL^2} \left[1-\lhat\,f'\!\left(-12/\tL^2\right)\right]-\frac{\lhat}{6}\,f\!\left(-12/\tL^2\right)\,. 
\ee
Hence determining the HEE will amount to finding the extremal area surface and evaluating eq.~\reef{RyuTaka} with an additional overall coefficient of 
\begin{eqnarray}\label{alfr}
\alhat=1+\lhat\, f^{\prime}(\bar R) \,.
\end{eqnarray}
Hence with a corner in the boundary entangling surface, the expression for the HEE will be a trivial generalization of eq.~\reef{eek5} with
\begin{eqnarray}
\labell{seerfr}
 S_{f(R)}=\alhat\,\left[\frac{\tilde{L}^2}{2G} \frac{H}{\delta}-\qe(\Omega) \log \left(\frac{H}{\delta} \right)+ \mathcal{O}(1)\right]  \, .
\end{eqnarray}

However, we emphasize that the same overall factor \reef{alfr} will appear in front of the entanglement entropy for any entangling surface and, in particular, for the circle. Further, it can be shown that the planar black hole solution (\ref{bb}) to the (four-dimensional) Einstein equations will also be a solution of the $f(R)$ Lagrangian. Hence the thermal entropy, which is computed by evaluating the horizon entropy using the same Wald formula \reef{seerfr}, will produce the Einstein gravity result (\ref{tebb}), again up to an overall factor given precisely by $\alhat$.  Hence for this class of theories, the ratios $\kappa/c_0$ and $\kappa/\cs$ will match those in Einstein gravity, as given in eqs.~\reef{ratioE1} and \reef{ratioE2}, respectively. Note that these results apply even when the strength of the gravitational couplings is large, \ie the fact that these ratios do not change is not restricted to linear order in perturbative calculations.

In order to see what happens with the two-point function \reef{t2p} of the stress tensor, we can follow the steps of section (\ref{tt}) in order to find the linearized equations of motion for the massless spin-two graviton in the AdS$_4$ background. A remarkable fact about our previous linearized equations (\ref{eomg}) was that none of the theories considered except that with an $R_{\mu\nu}R^{\mu\nu}$ interaction produced terms involving higher-order derivatives acting on $h_{\mu\nu}$ after we imposed the transverse traceless gauge. That is, in general, these theories do produce fourth-order derivatives of $h_{\mu\nu}$ in the linearized equations, but nevertheless these contributions all vanish in the AdS$_4$ background, with the exception of the $\lambda_2$ term, after we set $\bar{\nabla}^{\mu}h_{\mu\nu}=0=h\equiv {\bar{g}}^{\mu\nu}h_{\mu\nu}$. As an illustrative exercise, we explicitly demonstrate how this works in the case of $f(R)$ gravity, where the same behavior is encountered. The full linearized equations arising from eq.~(\ref{fr}) read
\begin{equation}\labell{linfr}
R^L_{\mu\nu}-\frac{1}{2}\bar{g}_{\mu\nu}R^L+\left[\frac{6}{\tilde{L}^2}-\frac{3}{L^2}\right]h_{\mu\nu}+\lhat\,\mathcal{E}_{\mu\nu}=0\, , 
\end{equation}
where
\begin{equation}
\mathcal{E}_{\mu\nu}\equiv f^{\prime}(\bar{R}) R_{\mu\nu}^L -\frac{1}{2}f(\bar{R}) h_{\mu\nu}+f^{\prime\prime}(\bar{R}) \left[\bar{g}_{\mu\nu} \bar{\square} -\bar{\nabla}_{\mu}\bar{\nabla}_{\nu}-\frac{3}{\tilde{L}^2}\bar{g}_{\mu\nu} \right]\! R^L-\frac{1}{2}f^{\prime}(\bar{R})\bar{g}_{\mu\nu} R^L\, ,
\end{equation}
and where the linearized Ricci tensor and Ricci scalar can be written as
\begin{eqnarray}
R_{\mu\nu}^L&=&-\frac{4}{\tilde{L}^2}h_{\mu\nu}+\frac{1}{\tilde{L}^2}\bar{g}_{\mu\nu}h+\frac12\left( \bar{\nabla}_{\mu}\bar{\nabla}_{\sigma}h_{\nu}{}^{\sigma} +
\bar{\nabla}_{\nu}\bar{\nabla}_{\sigma}h_{\mu}{}^{\sigma}-\bar{\square}h_{\mu\nu}-\bar{\nabla}_{\mu}
\bar{\nabla}_{\nu}h\right)\, , 
\nonumber \\ \labell{bmper}
R^L&\equiv&\bar g^{\mu\nu}R_{\mu\nu}^L-h^{\mu\nu}\bar R_{\mu\nu}=\bar{\nabla}^{\mu}\bar{\nabla}^{\nu}h_{\mu\nu}-\bar{\square}h+\frac{3}{\tilde{L}^2}h\, .
\end{eqnarray}
As we can see, these equations involve fourth-order derivatives of the perturbation and its trace. However, in the transverse traceless gauge, it is straightforward to see that $R^L$ vanishes and hence the fourth-order terms, which all appear in ${\cal E}_{\mu\nu}$, also vanish. The equations \reef{linfr} are then notably simpler and after some massaging,\footnote{In particular, one needs to use eq.~\reef{racket} in order to obtain eq.~(\ref{alfaf}).} they yield the result:
\begin{equation} \labell{alfaf}
- \frac{\alhat}{2}\left[\bar{\Box}+\frac{2}{\tilde{L}^2}\right]h_{\mu\nu}=\alhat\, G^L_{\mu\nu}=0\, ,
\end{equation}
where $G^L_{\mu\nu}$ is again the linearized Einstein tensor in this gauge, as in eq.~(\ref{lineaR}), and $\alhat$ is defined in eq.~(\ref{alfr}). Hence with this exercise, we see all the fourth-order terms explicitly disappear from the linearized equations. Further, we can see the same overall constant \reef{alfr} will appear here in $\ctt$, as appeared in the corner contribution above. Therefore the ratio $\kappa/\ctt$ is again unchanged from the Einstein value \reef{ratioE3} in the holographic CFT's dual to $f(R)$ gravity. 

When we impose the transverse traceless gauge, we are implicitly eliminating any new degrees of freedom and focusing entirely on the physical spin-two graviton. However, we know that $f(R)$ gravity introduces an additional scalar degree of freedom \cite{revfr} -- in particular, the trace of the metric perturbation becomes a propagating massive scalar field. Hence it is interesting exercise to relax the transverse traceless condition to see the extra scalar emerge. In order to find its equation, we can take the trace of the full linearized equations (\ref{linfr}) without any gauge fixing. The result is
\begin{eqnarray}
&&-\left[\alhat+\frac{12\lhat}{\tilde{L}^2}f^{\prime\prime}(\bar{R}) \right]\,\left[\bar{\nabla}^{\mu}\bar{\nabla}^{\nu}h_{\mu\nu}+\frac{3h}{\tilde{L}^2}\right] \\ \notag
&&\quad+\bar{\square}h\left[\alhat+\frac{21\lhat}{\tilde{L}^2}f^{\prime\prime}(\bar{R}) \right]
+3\lhat\, f^{\prime\prime}(\bar{R})\left[\bar{\square}\bar{\nabla}^{\mu}\bar{\nabla}^{\nu}h_{\mu\nu}-\bar{\square}^2h \right]=0\, . 
\end{eqnarray}
At this point, it is convenient to choose the gauge condition,
\be\labell{gagcon}
\bar{\nabla}^{\mu}h_{\mu\nu}=\bar{\nabla}_{\nu}h\,,
\ee
because this choice actually eliminates the fourth-order derivatives in the previous equation. The remaining second-order equation then simplifies to
\begin{eqnarray}\label{eqh}
3\lhat\, f^{\prime\prime}(\bar{R}) \,\bar{\square}h-\left[\alhat+\frac{12\lhat}{\tilde{L}^2}f^{\prime\prime}(\bar{R}) \right]h  =0\, ,
\end{eqnarray}
which corresponds to the equation of motion for a massive scalar field, as long as $f^{\prime\prime}(\bar{R})\neq 0$. That is, the trace $h$ has become a dynamical degree of freedom in this case. On the other hand, if $f^{\prime\prime}(\bar{R})= 0$ (\eg as in Einstein gravity), the above equation is not dynamical and would simply impose the tracelessness condition $h=0$. Hence the spin-two graviton would be the only propagating degree of freedome in this case.

We should also consider the traceless part of the metric perturbation, \ie 
\be\labell{notrace}
\hat h_{\mu\nu}=h_{\mu\nu}-\frac14\bar g_{\mu\nu}\,h\,.
\ee
with the gauge condition \reef{gagcon}. Combining this choice of gauge with eqs.~\reef{linfr} and \reef{eqh}, we find
\be \label{traceless3}
- \frac{\alhat}{2}\left[\bar{\Box}+\frac{2}{\tilde{L}^2}\right]\hat h_{\mu\nu}=-\frac{1}{2}\left[\alhat-\frac{6\lhat}{\tilde{L}^2}f^{\prime\prime}(\bar{R}) \right]\left[\bar{\nabla}_{\mu}\bar{\nabla}_{\nu}h-\frac{1}{4}\bar{g}_{\mu\nu}\bar{\square} h \right]\, .
\ee
which is almost the expected equation  for the massless spin-2 field $\hat{h}_{\mu\nu}$, \ie eq.~\reef{alfaf}, except that it includes a source term that is linear in the trace $h$. We can nevertheless define a new traceless tensor satisfying an equation with the same form as eq.~(\ref{alfaf}). This is given by\footnote{The procedure followed here for identifying the physical spin-two field closely follows \cite{Smolic:2013gz}, where the analysis was done for curvature-squared gravities.}
\be \label{ttensor}
t_{\mu\nu}\equiv \hat{h}_{\mu\nu}-\left[\frac{3\lhat f^{\prime\prime}(\bar{R})}{\alhat} \right] \left[\bar{\nabla}_{\mu}\bar{\nabla}_{\nu}h-\frac{1}{4}\bar{g}_{\mu\nu}\bar{\square} h \right]\, .
\ee
Indeed, using (\ref{notrace}) and (\ref{traceless3}), it can be shown that this tensor satisfies
\be
- \frac{\alhat}{2}\left[\bar{\Box}+\frac{2}{\tilde{L}^2}\right]t_{\mu\nu}=0\, .
\ee
Hence $t_{\mu\nu}$ represents the physical massless spin-2 graviton coupling to the holographic stress tensor. Note that the redefinition in eq.~(\ref{ttensor}) is trivial whenever $ f^{\prime\prime}(\bar{R})=0$ (as in Einstein gravity), so in that case, the traceless perturbation $\hat{h}_{\mu\nu}$ already corresponds to the massless spin-two mode.



It is interesting to consider the scalar degree of freedom more explicitly. Hence let us consider the case of  $R^2$ gravity, for which we write $\lhat\,f(R)= \lambda_1 L^2 R^2$. Hence we have $\lhat\, f^{\prime}(\bar{R})=2\lambda_1 L^2\bar{R}=-24\lambda_1 L^2/\tilde{L}^2$ and $\lhat\, f^{\prime\prime}(\bar{R})=2\lambda_1 L^2$. Further, as noted in section \ref{cs}, the solution of eq.~\reef{racket} is simply $\tL={L}$. Combining these expressions in  eq.~(\ref{eqh}) then yields
\begin{eqnarray}\label{eqhr2}
\left[\lambda_1\,\bar{\square}-\frac{1}{6L^2} \right] h =0\, ,
\end{eqnarray}
and hence $h$ obeys the standard equation of motion for a free scalar with mass: $m^2=1/(6\lambda_1 L^2)$. Using the standard holographic dictionary \cite{Maldacena:1997re,Witten:1998qj,Gubser:1998bc}, $h$ is dual to a scalar operator in the three-dimensional boundary CFT with 
\be\labell{pout}
\Delta=\frac{3}2+\sqrt{\frac{9}4+\frac{1}{6\lambda_1}}\,.
\ee
Hence if $\lambda_1$ is small and positive, $h$ corresponds to a highly irrelevant operator with $\Delta\simeq1/\sqrt{6\lambda_1}$ and with positive norm. If $\lambda_1$ is small and negative, $\Delta$ becomes imaginary indicating that the standard AdS/CFT dictionary is breaking down. In this limit, $h$ is a ghost-like scalar with a tachyonic mass which exceeds the Breitenloner-Freedman bound \cite{BF1,BF2}. Hence the bulk theory would be inherently unstable if we tried to interpret the corresponding $R^2$ gravity as a complete theory rather than as an effective low energy theory. Further, for $R^2$ gravity, eqs.~(\ref{traceless3}) and (\ref{ttensor}) reduce  to
\be \label{traceless3x}
\frac{1}{2}\left[\bar{\nabla}_{\mu}\bar{\nabla}_{\nu}h-\frac{1}{4}\bar{g}_{\mu\nu}\bar{\square} h \right]\left[1-36\lambda_1\right]- \frac{1-24\lambda_1}{2}\left[\bar{\Box}+\frac{2}{\tilde{L}^2}\right]\hat h_{\mu\nu}=0\, ,
\ee
and
\be \label{ttensorx}
t_{\mu\nu}\equiv \hat{h}_{\mu\nu}-\left[\frac{6\lambda_1}{1-24\lambda_1} \right] \left[\bar{\nabla}_{\mu}\bar{\nabla}_{\nu}h-\frac{1}{4}\bar{g}_{\mu\nu}\bar{\square} h \right]\, ,
\ee
which is in agreement with the results obtained in \cite{Lu:2011zk,Smolic:2013gz}.


\section{Free field results for $\sigma$}\label{integrals}

In \cite{Casini:2009sr,Casini:2006hu,Casini:2008as}, the first fourteen coefficients in Taylor expansion of $\q(\Omega)$ around $\Omega=\pi$ were computed for a free scalar and a free Dirac fermion using quantum field theory techniques. Of course,  in eq.~\reef{sigmA}, we identified the first coefficient in this expansion as the charge $\sigma$, \ie $\q(\Omega)=\sigma\, (\Omega-\pi)^2+\mathcal{O}\left((\Omega-\pi)^4\right)$. In the discussion section, we saw that the ratio of $\sigma$ with the central charge $\ctt$ in the two-point function of the stress tensor (\ref{t2p}) appears to satisfy a universal relation of the form
\beq \labell{sct}
\frac{\sigma}{C_{\ssc T}}=\frac{\pi^2}{24}\, .
\eeq
This analytic result was obtained using holographic techniques in the previous sections, and holds for all the higher-order gravity theories which were considered in this paper. In addition, the same ratio was obtained for the free scalar and the free fermion with an accuracy better than $0.0003\%$ using the results for $\sigma$ in \cite{Casini:2009sr} and for $\ctt$ in \cite{Osborn:1993cr}. The values for  $\ctt{}_{\rm \!, scalar}$ and $\ctt{}_{\rm \!, fermion}$ are exact --- see eq.~(\ref{ctsf}). However, the values of $\sigma_{\rm scalar}$ and $\sigma_{\rm fermion}$ can only be computed numerically. In particular, they can be obtained by evaluation of the following (monstruous) integrals\footnote{We wish to thank Horacio Casini for providing these integrals, which are the ones originally used in \cite{Casini:2009sr}.}

\begin{eqnarray}\label{fun1}
\sigma_{\rm scalar}&=&-2 \pi\,\int_{1/2}^{+\infty} dm \int_{0}^{+\infty}db\  \mu\, H\,a (1-a) \,m\,  \text{sech}^2(\pi  b)\, ,\\ \label{fun2}
\sigma_{\rm fermion}&=&-4\pi\int_{1/2}^{+\infty} dm \int_{0}^{+\infty}db\, \left[\mu \,H\, a(1-a)-\frac{F}{4\pi}\right]\,m \, \text{cosech}^2(\pi b)\, ,
\end{eqnarray}
where 
\begin{eqnarray}
\notag
H&\equiv&-\frac{c}{2h} X_1T-\frac{1}{2c} X_2T+\frac{1}{16 \pi  a (a-1) }\, , \\ \notag
h&\equiv&\frac{2 \left(a(a-1) +m^2\right) \sin ^2(\pi  a)}{m^2 \left(\cos (2 \pi  a)+\cos \left(\pi  \sqrt{1-4 m^2}\right)\right)}\, ,\\ \notag
c&\equiv&\frac{ 2^{2 a-1}\pi a (1-a) \, \sec \left(\frac{\pi}{2} \left(2 a+\sqrt{1-4 m^2}\right)\right)\,\Gamma\! \left(\frac{3}{2}-a+\frac12\sqrt{1-4
   m^2}\right)}{m\, \Gamma (2-a)^2\, \Gamma\! \left(a-\frac12+\frac{1}{2} \sqrt{1-4 m^2}\right)}\, ,\\ \notag
X_1&\equiv&-\frac{ \Gamma (-a) \left[\pi  \sinh \left(\frac{\pi  \mu }{2}\right)+ i \cosh \left(\frac{\pi  \mu }{2}\right) \left(\psi^{(0)}\!\left(a+\frac{i \mu }{2}+\frac{1}{2}\right)-\psi ^{(0)}\!\left(a-\frac{i \mu }{2}+\frac{1}{2}\right)\right)\right]}{2^{2 a+1}\mu \, \Gamma (a+1)\, \Gamma\!
   \left(-a-\frac{i \mu }{2}+\frac{1}{2}\right)\, \Gamma\! \left(-a+\frac{i \mu }{2}+\frac{1}{2}\right) (\cos (2 \pi  a)+\cosh (\pi  \mu ))}\,,\\ 
   X_2&\equiv&{\rm ``}X_1{\rm "}\ \,\, {\rm with } \,\,\, a\,\, \, {\rm replaced\,\,\, by }\, \,\, (1-a),\, \\ \notag
T&\equiv& \sqrt{h(a^2-a+(h+1)m^2)}\, ,
\end{eqnarray}
\begin{eqnarray}
F&\equiv&-\frac{F_1}{F_2}\, ,\\ \notag
F_1&\equiv & a^2 \left(8 \pi  c^2 \left(m^2+1\right) X_1 T+8 \pi  h
   \left(m^2+1\right)X_2T-c h\right)-16 \pi  a^3T \left(c^2
   X_1+h X_2\right)\\ \notag&+&a \left(-8 \pi  c^2 m^2 X_1T-8 \pi h m^2 X_2T+c h\right)+8 \pi  a^4 T \left(c^2 X_1+h X_2\right)-c h (h+1) m^2\, ,\\ \notag
F_2&\equiv &\frac{8 c\, h \left(a^2-a+m^2\right)^2}{(2 a-1) \mu}\, ,\\ \notag
\mu&\equiv& \sqrt{4m^2-1}\, ,\\ \notag
%
a&\equiv&\left\lbrace \begin{array}{cll}
 i\,b  +\frac{1}{2}&&\text{for the scalar}\, ,\\ 
i\,b && \text{for the fermion}\, , \end{array}\right.
\end{eqnarray}
and $\psi ^{(0)}$ denotes the usual digamma function  \ie $\psi ^{(0)}(z)=\frac{d\ }{dz}\log\Gamma(z)$.
Notice that eqs.~(\ref{fun1}) and (\ref{fun2}) look very different and without further insight, there is no reason to believe that these integrals should produce simple fractions or that the results should agree up to a factor $2$.

We can compute these integrals (\ref{fun1}) and (\ref{fun2}) numerically with arbitrary precision (although, of course, the computation time increases considerably as we increase the precision). Our results indicate that both eqs.~(\ref{fun1}) and (\ref{fun2}) exactly produce the results predicted assuming that $\sigma/\ctt$ is given by the universal constant in eq.~(\ref{sct}), \ie
\be
\sigma_{\rm scalar}=\frac{1}{256}=0.00390625\,,\qquad
\sigma_{\rm fermion}=\frac{1}{128}=0.0078125\,.
\ee
We  have verified this result  to a precision of approximately one part in $10^{12}$. In particular, we find
\be
\sigma_{\rm scalar}=0.00390625000000(5)\,,\qquad
\sigma_{\rm fermion}=0.00781250000000(7)\,,
\ee
where the numbers in brackets are out of the range of accuracy of our computation.

\end{document}